\documentclass[graybox]{svmult}

\usepackage{type1cm}        % activate if the above 3 fonts are
\usepackage{makeidx}         % allows index generation
\usepackage{graphicx}        % standard LaTeX graphics tool
\usepackage{multicol}        % used for the two-column index
\usepackage[bottom]{footmisc}% places footnotes at page bottom

\usepackage{newtxtext}       % 
\usepackage[varvw]{newtxmath}       % selects Times Roman as basic font
\usepackage[usestackEOL]{stackengine}

\newlength{\tempdima}
\newcommand{\rownameform}[1]% #1 = text
{\rotatebox{90}{\makebox[\tempdima][c]{\hspace{4cm}{{#1}}}}}

\usepackage{amsmath}
\usepackage{bm}
\usepackage{siunitx}

\usepackage{booktabs}
\usepackage{array}
\usepackage{mathtools}% http://ctan.org/pkg/mathtools
\usepackage{multicol}
\usepackage{multirow}

\usepackage{automated_macros}

\graphicspath{{figures/}}
\makeindex             % used for the subject index
\begin{document}

\title*{Physics-Informed Neural Networks for Solving Contact Problems in Three Dimensions}
\author{Tarik Sahin\orcidID{0000-0002-4134-3726},\\
        Daniel Wolff\orcidID{0000-0001-5767-7803} and\\
        Alexander Popp\orcidID{0000-0002-8820-466X}}
\institute{Tarik Sahin \at University of the Bundeswehr Munich, Institute for Mathematics and Computer-Based Simulation (IMCS), \email{tarik.sahin@unibw.de}
\and Daniel Wolff \at University of the Bundeswehr Munich, Institute for Mathematics and Computer-Based Simulation (IMCS) \email{d.wolff@unibw.de}
\and Alexander Popp \at University of the Bundeswehr Munich, Institute for Mathematics and Computer-Based Simulation (IMCS) \email{alexander.popp@unibw.de}}
\maketitle

%\abstract*{}
\abstract{
This paper explores the application of physics-informed neural networks (PINNs) to tackle forward 
problems in 3D contact mechanics, focusing on small deformation elasticity. We utilize a mixed-variable formulation, 
enhanced with output transformations, to enforce Dirichlet and Neumann boundary conditions as hard 
constraints. The inherent inequality constraints in contact mechanics, particularly the Karush-Kuhn-Tucker (KKT) 
conditions, are addressed as soft constraints by integrating them into the network's loss function. To enforce the 
KKT conditions, we leverage the nonlinear complementarity problem (NCP) approach, specifically using the 
Fischer-Burmeister function, which is known for its advantageous properties in optimization.
We investigate two benchmark examples of PINNs in 3D contact mechanics: a single contact patch test and the Hertzian contact problem. 
% For the Hertzian contact problem, we examine both forward and inverse cases. In the forward case, a PINN acts as PDE 
% solver to simulate the contact mechanics problem and identify the unknown displacement and stress fields. 
% Additionally, we demonstrate that combining the Adam and L-BFGS-B optimizers leads to improved accuracy 
% and faster convergence during training.
}

% \abstract{This paper investigates the application of physics-informed neural networks (PINNs) to solve forward and 
% inverse problems in 3D contact mechanics for small deformation elasticity. We adopt a mixed-variable formulation for 
% PINNs, incorporating output transformations to enforce Dirichlet and Neumann boundary conditions as hard constraints. 
% The inequality constraints of contact mechanics, particularly the Karush-Kuhn-Tucker (KKT) conditions, are handled as 
% soft constraints by embedding them into the network's loss function. In this work, to enforce the KKT conditions, we 
% employ a nonlinear complementarity problem (NCP) function, namely the Fischer-Burmeister function, which offers 
% favorable optimization properties. The two investigated examples of PINNs for 3D contact mechanics are single contact patch test 
% and the Hertzian contact problem. Using the Hertzian contact problem, we present two cases: a forward model where PINNs 
% function as pure partial differential equation (PDE) solvers and an inverse model for parameter identification. 
% We highlight the significance of choosing appropriate hyperparameters, such as loss weights, and demonstrate the 
% benefits of combining the \textit{Adam} and \textit{L-BFGS-B} optimizers to improve both accuracy and training 
% efficiency.}

%%%%%%%%%%%%%%%%%%%%%%%%%%%%%%%%%%%%%%%%%%%%%%%%%%%%%%%%%%%%%%%%%%%%%%%%%%%%%%%%%%%%%%%%%%%%%%%%%%%%%%%%%%%%%%%%%%%%%%%%%%%%%%
\section{Introduction}
\label{sec:intro}
% Introduction
% Machine learning methods often require substantial amounts of simulation or experimental data, which can be difficult 
% to obtain due to the complexity of simulations and the high costs associated with experiments. Data scarcity can also 
% lead to poor performance in data-driven models, particularly in scenarios involving real-world observations where data 
% may be noisy or incorrectly labeled, as these models lack a physics-based feedback mechanism to validate their 
% predictions. To address this challenge, physics-informed neural networks (PINNs) have been introduced 
% \cite{raissi2019physics}.

Machine learning methods often demand large datasets, which are challenging to obtain 
in engineering due to costly experiments, the unavailability of certain measurements 
(e.g., inaccessible components or the absence of suitable sensors), or long simulation 
runtimes. When real-world data is available, it is frequently noisy or mislabeled. 
These limitations can undermine the performance of classical data-driven models, 
as they lack a physics-based mechanism to validate predictions. 
Physics-informed neural networks (PINNs) have been introduced to address these 
challenges \cite{raissi2019physics}.

% PINNs combine boundary or initial boundary value problems with measurement data within the neural network's loss 
% function, helping mitigate issues related to insufficient data and the black-box nature of purely data-driven approaches. 
% For forward problems, PINNs can act as partial differential equation (PDE) solvers, even for domains with irregular 
% geometries. This is possible because PINNs rely on automatic differentiation, eliminating the need for connectivity 
% between sampling points, making them a mesh-free method \cite{arend2022truly}. Additionally, PINNs offer the advantage of overcoming the 
% curse of dimensionality when approximating functions in higher dimensions \cite{poggio2017and}. 
% PINNs can be also used for thermo-mechanical coupling problems using mixed formulation \cite{haradni}. 
% Furthermore, PINNs are well-suited for 
% solving inverse problems due to their ability to easily incorporate measurement data \cite{sahin2024hybrid}.

Using automatic differentiation, PINNs integrate physical knowledge 
(e.g., partial differential equations (PDEs)) into the neural network's loss function, 
introducing physics-based regularization that mitigates the black-box nature of purely 
data-driven approaches. For forward problems, PINNs serve as mesh-free PDE solvers, 
requiring only sampling points from the (arbitrary, potentially geometrically intricate) 
domain \cite{arend2022truly}. They also overcome the curse of dimensionality when approximating high-dimensional 
functions \cite{poggio2017and} and excel in solving inverse problems by seamlessly incorporating measurement data 
\cite{sahin2024hybrid}.
PINNs can be also used for thermo-mechanical coupling problems using a mixed formulation 
\cite{haradni}.

In this work, we extend PINNs to 3D contact mechanics, 
building on previous 2D examples \cite{sahin2024}. The employed PINNs are developed based on  
a mixed-variable formulation inspired by the Hellinger-Reissner principle, where both 
displacement and stress fields are network outputs. Contact mechanics involves Karush-Kuhn-Tucker (KKT) type constraints, 
commonly referred to as Hertz-Signorini-Moreau conditions, which we incorporate as soft constraints using the Fischer-Burmeister nonlinear 
complementarity problem (NCP) function. This approach efficiently handles inequality constraints in contact mechanics. 
We validate the method with two examples assuming small deformation elasticity for simplicity:: (1) a single contact patch test where all points are in contact and (2) the Hertzian 
contact problem, where the contact area is detected as part of the solution. 
% Additionally, we apply PINNs in two ways 
% for the Hertzian problem: first as a forward solver to compare with finite element simulations, and second for inverse 
% problems, using FEM data to identify the external load.

%%%%%%%%%%%%%%%%%%%%%%%%%%%%%%%%%%%%%%%%%%%%%%%%%%%%%%%%%%%%%%%%%%%%%%%%%%%%%%%%%%%%%%%%%%%%%%%%%%%%%%%%%%%%%%%%%%%%%%%%%%%%%%
\section{Problem Formulation: Contact Mechanics}
\label{sec:problem_formulation}
\begin{figure}
    \centering
    \includegraphics[scale=0.475]{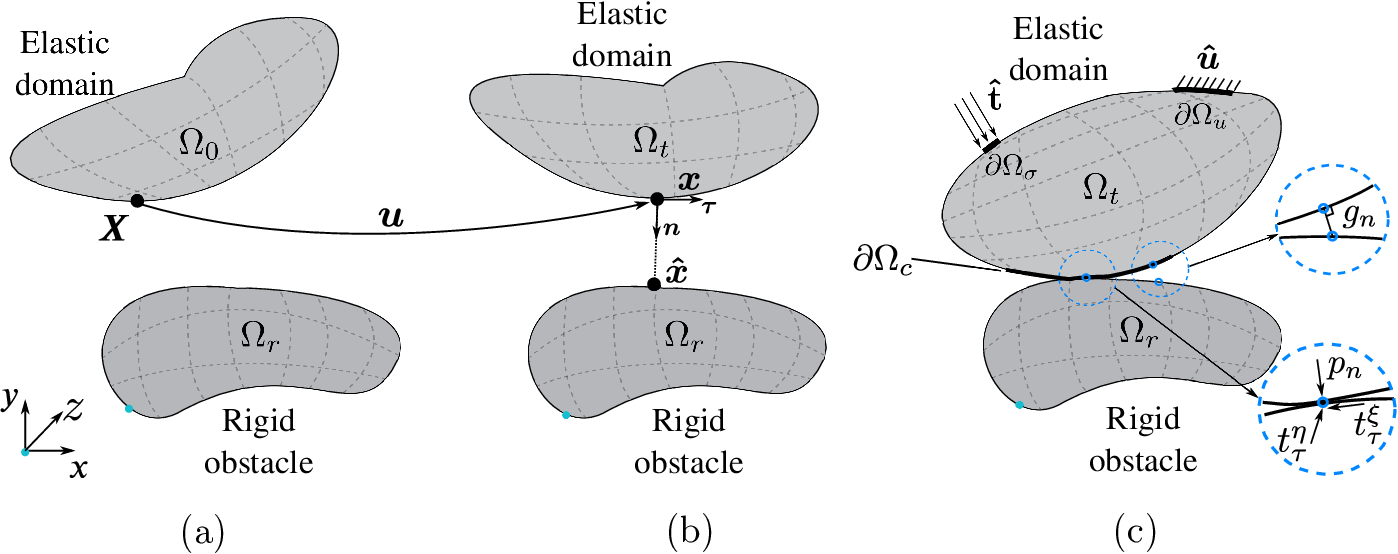}
    \caption{Contact problem between an elastic body and a rigid obstacle. 
    (a) Reference configuration, (b) current configuration, 
    (c) boundary conditions illustration, including the gap $g_n$, tangential tractions $t_{\tau}^{\xi}$, $t_{\tau}^{\eta}$ and contact pressure $p_n$.}
    \label{fig:contact_illus}
\end{figure}

We consider a 3D contact problem between an elastic body and a fixed rigid obstacle, as illustrated in 
Fig. \ref{fig:contact_illus}. In the reference configuration, the elastic body is denoted by $\Omega_{0}$, while in 
the current configuration, it is represented by $\Omega_{t}$. The rigid obstacle remains fixed in its configuration, $\Omega_{r}$. 
Fig. \ref{fig:contact_illus}c illustrates a scenario where the two bodies are in contact. The surface of the elastic 
body can be divided into three regions: the Dirichlet boundary $\partial \Omega_{u}$, where displacements are specified; 
the Neumann boundary $\partial \Omega_{\sigma}$, where tractions are applied; and the potential contact boundary 
$\partial \Omega_{c}$, where contact constraints may apply. The true contact surface is a subset of $\partial \Omega_{c}$ 
and is determined as part of the solution procedure.

We define the target boundary value problem (BVP) as follows:
\begin{align}
    \label{eq:lin_equi}
    \boldsymbol{\nabla \cdot } \boldsymbol{\sigma} + \bodyforce & = \mathbf{0} \hspace*{3mm} \text{in } \Omega_t, \\ 
    \boldsymbol{u} &= \boldsymbol{\hat u} \hspace*{2.6mm} \text{on } \partial \Omega_u, \\ 
    \boldsymbol{\sigma} \cdot \boldsymbol{n} &= \boldsymbol{\hat t} \quad \text{on } \partial \Omega_{\sigma}. %\nonumber
\end{align}
In this formulation, $\boldsymbol{\sigma}$ is the Cauchy stress tensor, and $\boldsymbol{u}$ is the displacement vector, 
which is the primary unknown. The vector $\bodyforce$ represents body forces acting on the domain, and $\boldsymbol{n}$ 
is the outward unit normal vector on the boundary. Displacements prescribed on the Dirichlet boundary 
$\partial \Omega_u$ are denoted by $\boldsymbol{\hat u}$, while tractions prescribed on the Neumann boundary 
$\partial \Omega_{\sigma}$ are represented by $\boldsymbol{\hat t}$.
% The terms BE, DBC, and NBC refer to the balance equation, Dirichlet boundary condition, and Neumann boundary condition, 
% respectively.

In addition to these equilibrium equations, we also account for the kinematic and constitutive equations that describe 
the behavior of the elastic material. For simplicity, linear kinematics and linear-elastic material behavior will be assumed in the numerical examples section, given by:
\begin{align}
    \label{eq:compatibility}
    \bm{\varepsilon} &= \frac{1}{2}(\boldsymbol{\nabla}\boldsymbol{u}+\boldsymbol{\nabla}\boldsymbol{u}^T), \\ 
    \label{eq:hooke}
    \boldsymbol{\sigma} &= \mathbb{C} : \bm{\varepsilon}.
\end{align}
Here, $\bm{\varepsilon}$ is the infinitesimal strain tensor, and $\mathbb{C}$ is the fourth-order elasticity tensor.

For the case of linear isotropic elasticity, the constitutive equation can be simplified using Hooke’s law:
\begin{equation}
    \boldsymbol{\sigma} = \lambda \tr(\bm{\varepsilon})\mathbf{I} + 2\mu \bm{\varepsilon},
\end{equation}
where $\lambda$ and $\mu$ are the Lamé parameters, $\tr(\cdot)$ denotes the trace operator (sum of diagonal components), 
and $\mathbf{I}$ is the identity tensor.

The displacement vector $\boldsymbol{u}$ of the elastic body is determined by describing the motion of a point in the reference 
configuration $\boldsymbol{X}\in \Omega_0$ to a point in the current configuration $\boldsymbol{x} \in \Omega_t$, as shown in 
Figs. \ref{fig:contact_illus}a and \ref{fig:contact_illus}b. This relation is expressed as follows:
\begin{equation}
    \label{eq:displacementvector}
    \boldsymbol{u} = \boldsymbol{x} - \boldsymbol{X}.
\end{equation}

The gap function $g_n$, which measures the separation between the elastic and rigid bodies in the current 
configuration, is defined as:
\begin{equation}
    g_n = -\boldsymbol{n}\cdot(\boldsymbol{x}-\boldsymbol{\hat x}).
\end{equation}
Here, $\boldsymbol{\hat x}$ denotes the closest point projection of $\boldsymbol{x}$ onto the surface of $\Omega_r$ 
(see Fig. \ref{fig:contact_illus}b). Since all contact constraints are enforced in the current configuration, 
the contact pressure $p_n$, and tangential traction components $t_{\tau}^{\xi}$ and $t_{\tau}^{\eta}$ and are obtained 
by decomposing the contact traction 
vector $\boldsymbol{t}_c$ via traction vector decomposition (TVD) as follows:
\begin{equation}
    \label{eq:contact_traction1}
    \boldsymbol{t}_c = p_n \boldsymbol{n} + t_{\tau}^{\xi} \boldsymbol{\tau}^{\xi} + t_{\tau}^{\eta} \boldsymbol{\tau}^{\eta}, 
    \quad p_n = \boldsymbol{t}_c \cdot \boldsymbol{n}, 
    \quad t_{\tau}^{\xi} = \boldsymbol{t}_c \cdot \boldsymbol{\tau}^{\xi},
    \quad t_{\tau}^{\eta} = \boldsymbol{t}_c \cdot \boldsymbol{\tau}^{\eta},
\end{equation}
where
\begin{equation}
    \label{eq:contact_traction2}
    \boldsymbol{t}_c = \boldsymbol{\sigma} \cdot \boldsymbol{n} \quad\quad \text{on } \partial \Omega_c. 
\end{equation}

For a frictionless contact problem, the classical Karush-Kuhn-Tucker (KKT) type constraints (Eq. \ref{eq:kkt}), also known as the 
Hertz-Signorini-Moreau (HSM) conditions, are employed along with the frictionless sliding conditions (Eq. \ref{eq:fst}):
\begin{align}
    g_n \geqslant 0, \quad p_n \leqslant 0 & , \quad p_n g_n = 0, \label{eq:kkt}\\%[0.25em]
    t_{\tau}^{\xi}=0, \quad & t_{\tau}^{\eta}=0. \label{eq:fst}
\end{align}
For further details on more advanced contact constitutive laws, including friction, the reader is referred 
to \cite{Popp2018a, wriggers2006computational}.
% \subsection{Mixed-variable Formulation: the Stress-Displacement Principle}
% \label{sec:mixed}

In this study, we adopt a mixed-variable approach, where both the displacement field $\boldsymbol{u}$ and the 
stress tensor $\boldsymbol{\sigma}$ are treated as primary variables for solving contact problems. 
An additional key requirement is the stress-to-stress coupling between the primary variable $\boldsymbol{\sigma}$ 
and the secondary variable $\boldsymbol{\sigma}^u$, defined by the relation
\begin{equation}
    \boldsymbol{\sigma} = \boldsymbol{\sigma}^u.  \label{eq:ss}
\end{equation}
The term $\boldsymbol{\sigma}^u$ is computed by inserting the kinematic equation (Eq. \ref{eq:compatibility}) 
into the constitutive law (Eq. \ref{eq:hooke}).
This condition ensures that the two master fields remain consistent and compatible throughout the solution process.
This method has been comprehensively discussed in the work by Sahin et al.~\cite{sahin2024}.
%%%%%%%%%%%%%%%%%%%%%%%%%%%%%%%%%%%%%%%%%%%%%%%%%%%%%%%%%%%%%%%%%%%%%%%%%%%%%%%%%%%%%%%%%%%%%%%%%%%%%%%%%%%%%%%%%%%%%%%%%%%%%%
\section{Extending Physics-Informed Neural Networks for 3D Contact Mechanics}
\label{sec:pinns_for_contact}

% \subsection{Generic PINNs Utilizing Output Transformation}
% \label{sec:generic_pinns}

\subsection{Loss Formulation in PINNs}
\label{sec:application_pinns_for_contact}
In quasi-static solid mechanics using PINNs, a fully-connected neural network (FNN) $\mathcal{N}$ with parameters $\boldsymbol{\theta}$ 
and spatial coordinates $\boldsymbol{x}$ to the displacement $\boldsymbol{u}$ and stress tensor $\boldsymbol{\sigma}$:
\begin{equation}
    \label{eq:approx_mixed}
    \pinnUVector = \mathcal{N}_{\boldsymbol{u}}(\boldsymbol{x};\boldsymbol{\theta})', \quad
    \pinnSVector = \mathcal{N}_{\boldsymbol{\sigma}}(\boldsymbol{x};\boldsymbol{\theta})'.
\end{equation}
Here, $(.)'$ denotes a user-defined output transformation and $(\tilde{.})$ indicates a predicted quantity.
It is important to note that, due to the conservation of angular momentum, 
$\boldsymbol{\sigma} = \boldsymbol{\sigma}^T$, only 6 stress components are predicted instead of 9.

The total loss $\pazocal{L}_E$ for the boundary value problem (without contact) in the mixed-variable formulation, with additional experimental data, is given by:
\begin{equation}
    \label{eq:pinn_elasticity}
    \pazocal{L}_E = \pazocal{L}_{\mathrm{PDEs}} + \pazocal{L}_{\mathrm{DBCs}} + \pazocal{L}_{\mathrm{NBCs}} + \pazocal{L}_{\mathrm{EXPs}},
\end{equation}
where:
\begin{align}
    \mathcal{L}_{\mathrm{PDEs}} &= \sum_{i_{n}=1}^{N_{m}} w_{i_{n}} \frac{1}{N_{rp}} \sum_{i_{rp}=1}^{N_{rp}} \Bigl[ \bigl[ \nabla \cdot \pinnSVector(\boldsymbol{x}^{i_{rp}}) + \bodyforce(\boldsymbol{x}^{i_{rp}}) \bigr]_{i_{n}} \Bigr]^2 \notag \\
                                 &+ \sum_{i_{n}=N_{m}+1}^{N_{n}} w_{i_{n}} \frac{1}{N_{rp}} \sum_{i_{rp}=1}^{N_{rp}} \Bigl[ \bigl[\pinnSVector(\boldsymbol{x}^{i_{rp}}) - \mathbb{C} : \tilde{\bm{\varepsilon}}(\boldsymbol{x}^{i_{rp}}) \bigr]_{i_{n}}\Bigr]^2, \\
    \mathcal{L}_{\mathrm{DBCs}} &= \sum_{i_{bc,D}=1}^{N_{bc,D}} w_{i_{bc,D}} \frac{1}{N_{bp,D}} \sum_{i_{bp,D}=1}^{N_{bp,D}} \Bigl[ \pinnUVector(\boldsymbol{x}^{i_{bp,D}}) - \boldsymbol{\hat u}(\boldsymbol{x}^{i_{bp,D}}) \Bigr]_{i_{bc,D}}^2, \\
    \mathcal{L}_{\mathrm{NBCs}} &= \sum_{i_{bc,N}=1}^{N_{bc,N}} w_{i_{bc,N}} \frac{1}{N_{bp,N}} \sum_{i_{bp,N}=1}^{N_{bp,N}} \Bigl[ \pinnSVector(\boldsymbol{x}^{i_{bp,N}}) \cdot \boldsymbol{n} - \mathbf{\hat t}(\boldsymbol{x}^{i_{bp,N}}) \Bigr]_{i_{bc,N}}^2, \\
    \mathcal{L}_{\mathrm{EXPs}} &= \sum_{i_{e,\boldsymbol{u}}=1}^{N_{e,\boldsymbol{u}}} w_{i_{e,\boldsymbol{u}}} \frac{1}{N_{ep,\boldsymbol{u}}} \sum_{i_{ep,\boldsymbol{u}}=1}^{N_{ep,\boldsymbol{u}}} \Bigl[ \pinnUVector(\boldsymbol{x}^{i_{ep,\boldsymbol{u}}}) - \boldsymbol{u^*}(\boldsymbol{x}^{i_{ep,\boldsymbol{u}}}) \Bigr]_{i_{e,\boldsymbol{u}}}^2 \notag \\
                             &+ \sum_{i_{e,\boldsymbol{\sigma}}=1}^{N_{e,\boldsymbol{\sigma}}} w_{i_{e,\boldsymbol{\sigma}}} \frac{1}{N_{ep,\boldsymbol{\sigma}}} \sum_{i_{ep,\boldsymbol{\sigma}}=1}^{N_{ep,\boldsymbol{\sigma}}} \Bigl[ \pinnSVector(\boldsymbol{x}^{i_{ep,\boldsymbol{\sigma}}}) - \boldsymbol{\sigma^*}(\boldsymbol{x}^{i_{ep,\boldsymbol{\sigma}}}) \Bigr]_{i_{e,\boldsymbol{\sigma}}}^2.
\end{align}
Here, $(.)^{\boldsymbol{*}}$ denotes measurements. The terms $\mathcal{L}_{\mathrm{DBCs}}$ and $\mathcal{L}_{\mathrm{NBCs}}$ represent the loss contributions from 
Dirichlet and Neumann boundary conditions, respectively, while $\mathcal{L}_{\mathrm{EXPs}}$ accounts for losses 
due to additional experimental data. In the mixed-variable formulation, $\mathcal{L}_{\mathrm{PDEs}}$ is composed of 
two parts: the balance equation (Eq. \ref{eq:lin_equi}) and the stress-to-stress coupling (Eq. \ref{eq:ss}). 
The index $N_{m}$ distinguishes between the loss weights for balance equation and stress-to-stress coupling.
The terms $\{w_{i_n}\}_{i_n=1}^{N_n}$, $\{w_{i_{bc}}\}_{i_{bc}=1}^{N_{bc}}$
and $\{w_{i_e}\}_{i_e=1}^{N_e}$ denote the loss weights for the individual components of $\pazocal{L}_{\mathrm{PDEs}}$,
$\pazocal{L}_{\mathrm{BCs}}$, $\pazocal{L}_{\mathrm{EXPs}}$, respectively. For a detailed explanation, we refer to \cite{sahin2024}. 
Since stress components are directly predicted by the network, Neumann BCs can be imposed as hard constraints using 
an output transformation. Moreover, only first-order derivatives of the network outputs are needed, as the governing 
equations in this formulation involve exclusively first-order derivatives.

For the problem formulation with contact, the total loss function $\pazocal{L}_{\mathrm{C}}$ is defined as:
\begin{equation}
    \label{eq:loss_contact}
    \pazocal{L}_{\mathrm{C}} = \pazocal{L}_{\mathrm{E}} + \pazocal{L}_{\mathrm{FS}} + \pazocal{L}_{\mathrm{KKT}},
\end{equation}
where $\pazocal{L}_{\mathrm{FS}}$ enforces the frictionless sliding conditions (Eq.~\ref{eq:fst}) on the contact surface $\partial \Omega_c$:
\begin{equation}
    \pazocal{L}_{\mathrm{FS}} = w_{\mathrm{fs,\xi}} | t_{\tau}^{\xi}|_{\partial \Omega_c} 
                                + w_{\mathrm{fs,\eta}} | t_{\tau}^{\eta}|_{\partial \Omega_c}.
\end{equation}
Here, $w_{\mathrm{fs,\xi}}$ and $w_{\mathrm{fs,\eta}}$ represent the loss weights corresponding
frictionless sliding conditions and 
$|\cdot|$ denotes the mean squared error (MSE), calculated as $\frac{1}{n}\sum_{i=1}^{n}(\cdot)^2$. 
While the evaluation of the normal gap $g_n$ is a topic of intense discussion in the context of discretization methods 
such as the finite element method \cite{wriggers2006computational},
it is consistently expressed by evaluating the orthogonal projection of 
the elastic body onto the rigid flat surface.
Additionally, $\pazocal{L}_{\mathrm{KKT}}$ comprises the Karush-Kuhn-Tucker conditions 
and will be introduced in the next section.

\subsection{Enforcing the Karush-Kuhn-Tucker Inequality Constraints: 
the \textit{Fischer-Burmeister} NCP Function}
\label{sec:enforcing_kkt}
Nonlinear complementarity problem (NCP) functions reformulate inequalities as equalities. 
A common NCP function is the Fischer-Burmeister function \cite{fischer1992}, defined as
\begin{equation}
    \label{eq:fischer_burmeister}
    \phi_{\mathrm{FB}}(a,b) := a + b - \sqrt{a^2 + b^2} = 0 \quad \Longleftrightarrow \quad a \geqslant 0, b \geqslant 0, ab = 0.
\end{equation}

By setting $a = \tilde{g}_n$ and $b = -\tilde{p}_n$, the Fischer-Burmeister function gives the KKT loss:
\begin{equation}
    \label{eq:fischer_burmeister_loss}
    \pazocal{L}_{\mathrm{KKT}} = w_{\mathrm{KKT}} \Bigl|\tilde{g}_n - \tilde{p}_n - \sqrt{{\tilde{g}_n}^2 + \tilde{p}_n^2}\Bigr|_{\partial \Omega_c},
\end{equation}
where, $w_{\mathrm{KKT}}$ denotes the loss weight. 
The Fischer-Burmeister function is ideal for mean squared error (MSE) loss calculations because $(\phi_{\mathrm{FB}})^2$ 
is continuously differentiable at $a = b = 0$. This method simplifies
optimization and parameter tuning by requiring only one loss weight instead of one per condition.

%%%%%%%%%%%%%%%%%%%%%%%%%%%%%%%%%%%%%%%%%%%%%%%%%%%%%%%%%%%%%%%%%%%%%%%%%%%%%%%%%%%%%%%%%%%%%%%%%%%%%%%%%%%%%%%%%%%%%%%%%%%%%%
\section{Numerical Examples}
\label{sec:numerical_examples}
The numerical examples share the following common settings. The PINN takes spatial coordinates 
$\boldsymbol{x} = (x,y,z)$ as inputs and maps them to transformed mixed-form outputs $(\pinnUVector,\pinnSVector)$. 
The networks are initialized using the \textit{Glorot uniform} initializer, and the activation function is set to 
\textit{tanh}. Training begins with the \textit{Adam} optimizer at a learning rate of $lr=0.001$ for 2000 epochs, 
followed by the \textit{L-BFGS-B} optimizer until the solver converges. 
Additionally, body forces are omitted. Unless otherwise specified, all loss weights are set to 1.

\subsection{Single Contact Patch Test}
\label{sec:single_patch}
% In the first example, we consider a single contact patch test as shown in Fig. \ref{fig:single_block_geometry}a. An 
% elastic cube is subjected to an external pressure on its top. The geometry is constrained in the $x$-direction at $x=0$
% and constrained in the $y$-direction at $y=0$. The analytical solution \cite{timoshenko1951theory} can easily derived as
% \begin{align*}
%     u_x = \nu \frac{p}{E}x,\quad  u_y = &-\frac{p}{E}y  \quad u_z = \nu \frac{p}{E}z, \\
%     \sigma_{yy} = -p, \quad \sigma_{xy}=0 & \quad \sigma_{yz}=0 \quad \sigma_{xz}=0.
% \end{align*}
% For our specific setup, the chosen parameters are $E=1.33$, $\nu=0.33$, $p=0.1$, $l=h=w=1$.

% We apply the following output transformation to enforce the DBCs and NBCs given in Fig. 
% \ref{fig:single_block_geometry}b as hard constraints 
% \begin{align*}
%     \pinnUComponent{x} = x \pinnUPredicter{x} \quad \pinnUComponent{z} = z \pinnUPredicter{z}, \quad 
%     & \pinnSComponent{xx}= (l-x)\pinnSPredicter{xx}, \\ 
%     \pinnSComponent{yy}= p +(h-y)\pinnSPredicter{yy}, \quad 
%     \pinnSComponent{zz}= (w-z)&\pinnSPredicter{zz}, \quad
%     \pinnSComponent{xy}= x(h-y)(l-x)\pinnSPredicter{xy},\\ 
%     \pinnSComponent{yz}= z(h-y)(w-z)\pinnSPredicter{yz},\quad
%     &\pinnSComponent{xz}= xz(l-x)(w-z)\pinnSPredicter{xz}.  
% \end{align*}
% This output transformation saves us from enforcing all DBCs and NBCs as soft constraints. On the other hands, 
% the contact constraints are enforced as soft constraints as explained in earlier sections. 

\begin{figure}[thbp]
    \centering
    \includegraphics[width=0.85\linewidth]{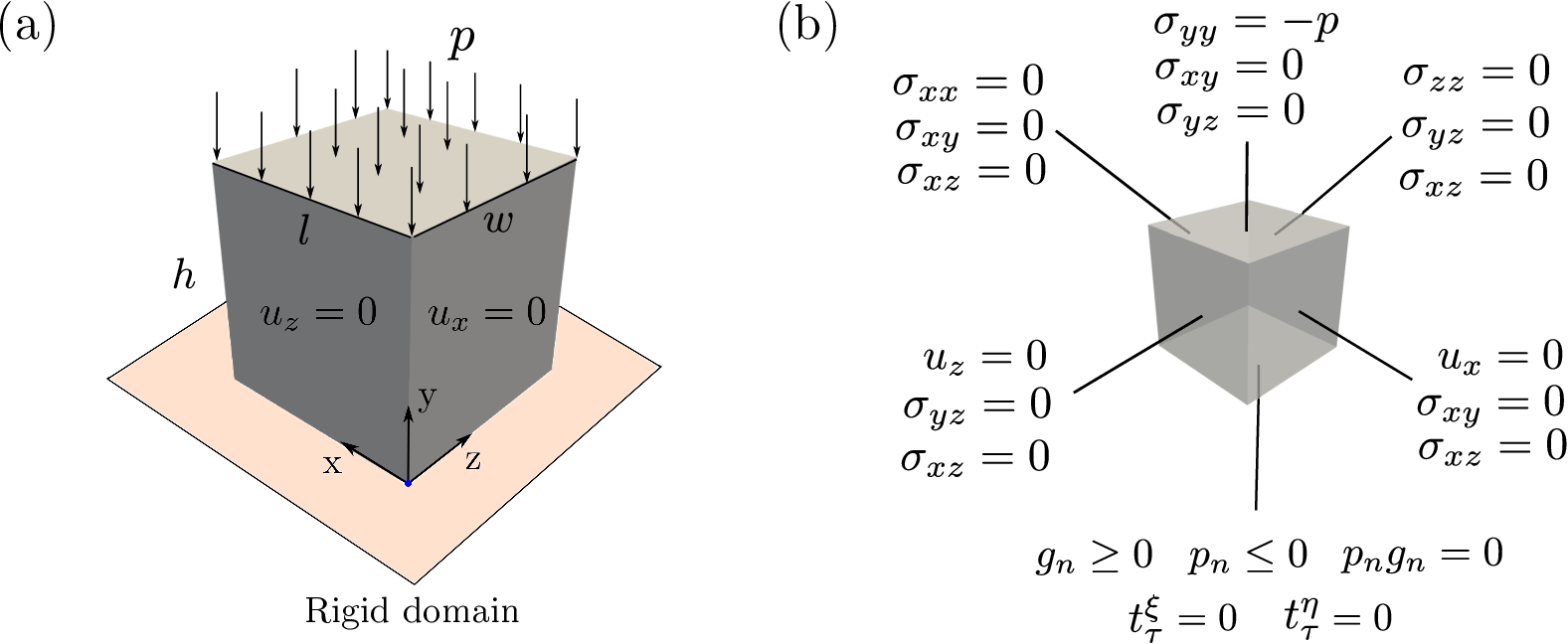}
    \caption{Single patch contact test problem:
    (a) geometry, constraints, and loading, 
    (b) all relevant boundary conditions.}
    \label{fig:single_block_geometry}
\end{figure}

In the first example, we examine a single patch contact test as depicted in Fig. \ref{fig:single_block_geometry}a. 
An elastic cube is subjected to external pressure on its top surface. The geometry is constrained in the $x$-direction at $x=0$ 
and in the $z$-direction at $z=0$. The analytical solution, as derived from \cite{timoshenko1951theory}, is given by:
\begin{equation*}
    u_x = \nu \frac{p}{E}x,\quad  u_y = -\frac{p}{E}y  \quad u_z = \nu \frac{p}{E}z, \quad \sigma_{yy} = -p.   
    % \sigma_{yy} = -p, \quad \sigma_{xy}=0 & \quad \sigma_{yz}=0 \quad \sigma_{xz}=0.
\end{equation*}
For this setup, the parameters are $E=1.33$, $\nu=0.33$, $p=0.1$, and $l=h=w=1$.

We use the following output transformation to impose the Dirichlet and Neumann boundary conditions (BCs) shown in 
Fig. \ref{fig:single_block_geometry}b as hard constraints:
\begin{align*}
    \pinnUComponent{x} = x \pinnUPredicter{x} \quad \pinnUComponent{z} = z \pinnUPredicter{z}, \quad 
    & \pinnSComponent{xx}= (l-x)\pinnSPredicter{xx}, \\ 
    \pinnSComponent{yy}= -p +(h-y)\pinnSPredicter{yy}, \quad 
    \pinnSComponent{zz}= (w-z)&\pinnSPredicter{zz}, \quad
    \pinnSComponent{xy}= x(h-y)(l-x)\pinnSPredicter{xy},\\ 
    \pinnSComponent{yz}= z(h-y)(w-z)\pinnSPredicter{yz},\quad
    &\pinnSComponent{xz}= xz(l-x)(w-z)\pinnSPredicter{xz}.  
\end{align*}

This transformation enforces the Dirichlet and Neumann BCs as hard constraints, while the contact conditions are treated 
as soft constraints, as described earlier. Additionally, the employed PINN is a fully connected neural network 
consisting of 5 hidden layers with 50 neurons each.

The comparison between the predicted and analytical solutions for the single contact patch test is shown in 
Fig. \ref{fig:single_block_geometry}. The results are only provided for the displacement components $u_x$ and $u_y$, 
as the behavior of $u_z$ is identical to that of $u_x$. 
Both predicted displacement fields exhibit a linear distribution that closely matches the analytical solution. 
While the contact constraints are imposed softly, the displacement component $u_y$ approaches zero near the contact 
surface, reflecting the expected behavior.
\begin{figure}
    \centering
    % Define figure size, horizontal spacing, and crop parameters
    \def\figsize{0.25}
    \def\hskiplocal{\hskip 0.5cm}
    \def\cropL{3.75cm}  % Crop from the left
    \def\cropB{6cm}  % Crop from the bottom
    \def\cropR{2cm}  % Crop from the right
    \def\cropT{3cm}  % Crop from the top
    \centering
    \def\arraystretch{1}% 
    \begin{tabular}{c@{\hskip 0.2cm}c@{\hskiplocal}c@{\hskiplocal}c}
        &\hspace{-1.2cm}\scalebox{0.9}{$u_x$}  
        &\hspace{-1.2cm}\scalebox{0.9}{$u_y$} 
        &\hspace{-1.2cm}\scalebox{0.9}{$\sigma_{yy}$} \\[1ex]
        \rownameform{\hspace{-2cm}\scalebox{0.9}{Prediction}}&
        \includegraphics[width=\figsize\linewidth, trim={{\cropL} {\cropB} {\cropR} {\cropT}}, clip]{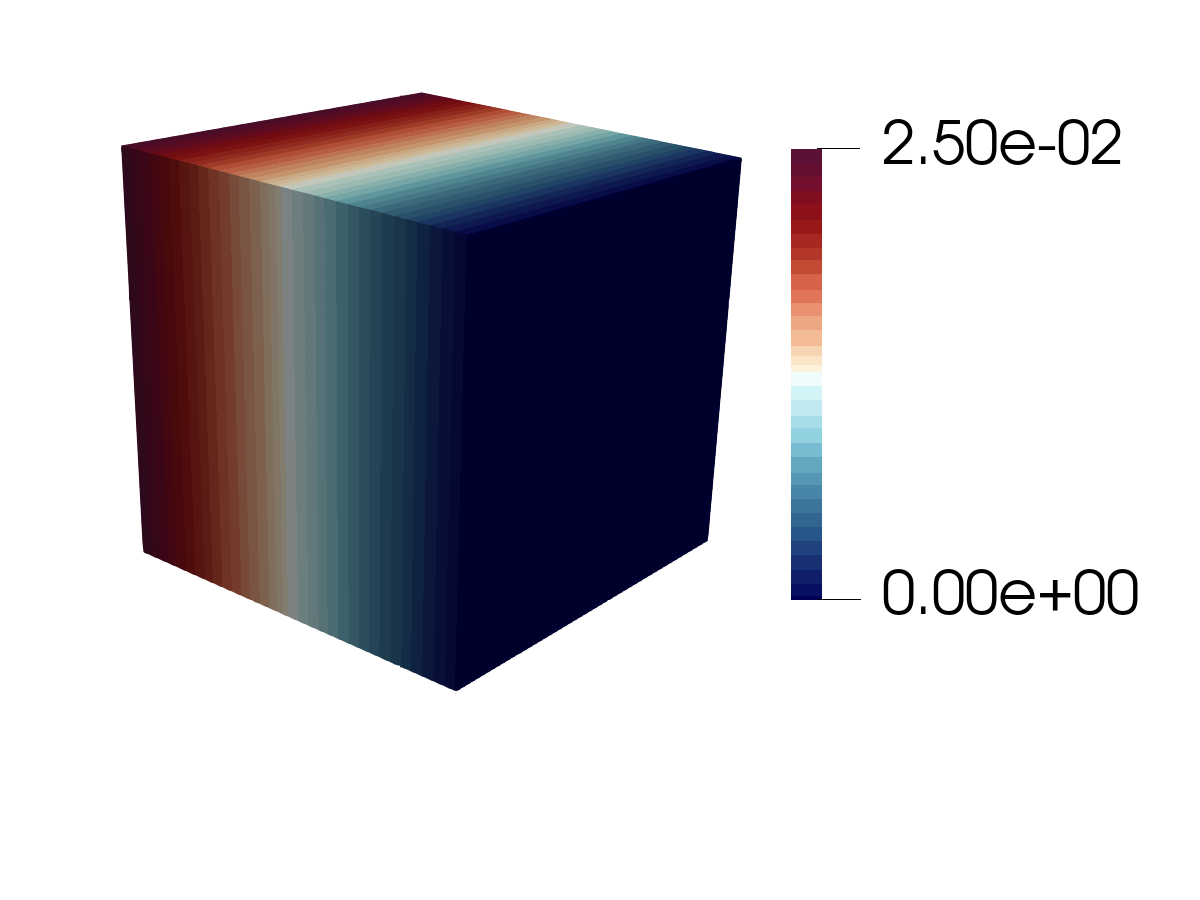}&
        \includegraphics[width=\figsize\linewidth, trim={{\cropL} {\cropB} {\cropR} {\cropT}}, clip]{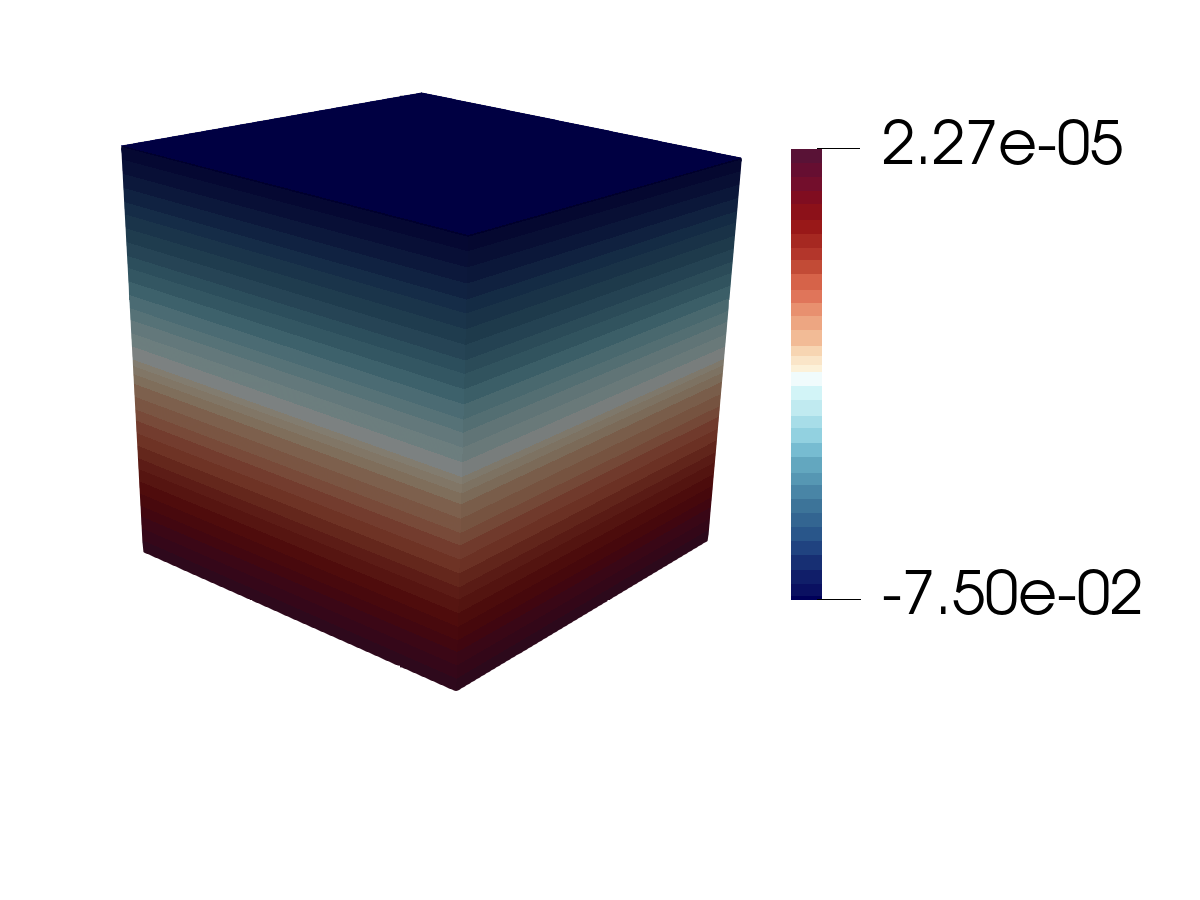}&
        \includegraphics[width=\figsize\linewidth, trim={{\cropL} {\cropB} {\cropR} {\cropT}}, clip]{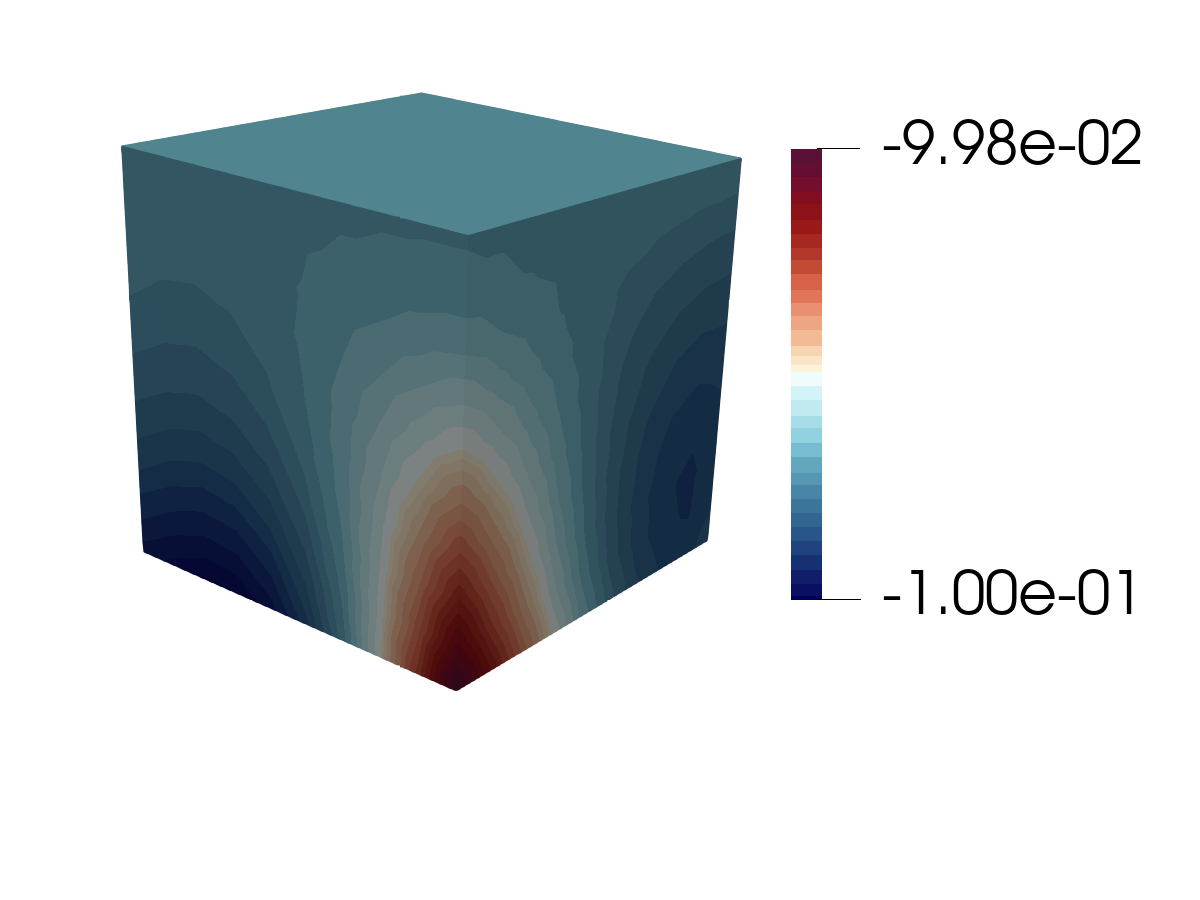}\\

        \rownameform{\hspace{-2cm}\scalebox{0.9}{Analytical}}&
        \includegraphics[width=\figsize\linewidth, trim={{\cropL} {\cropB} {\cropR} {\cropT}}, clip]{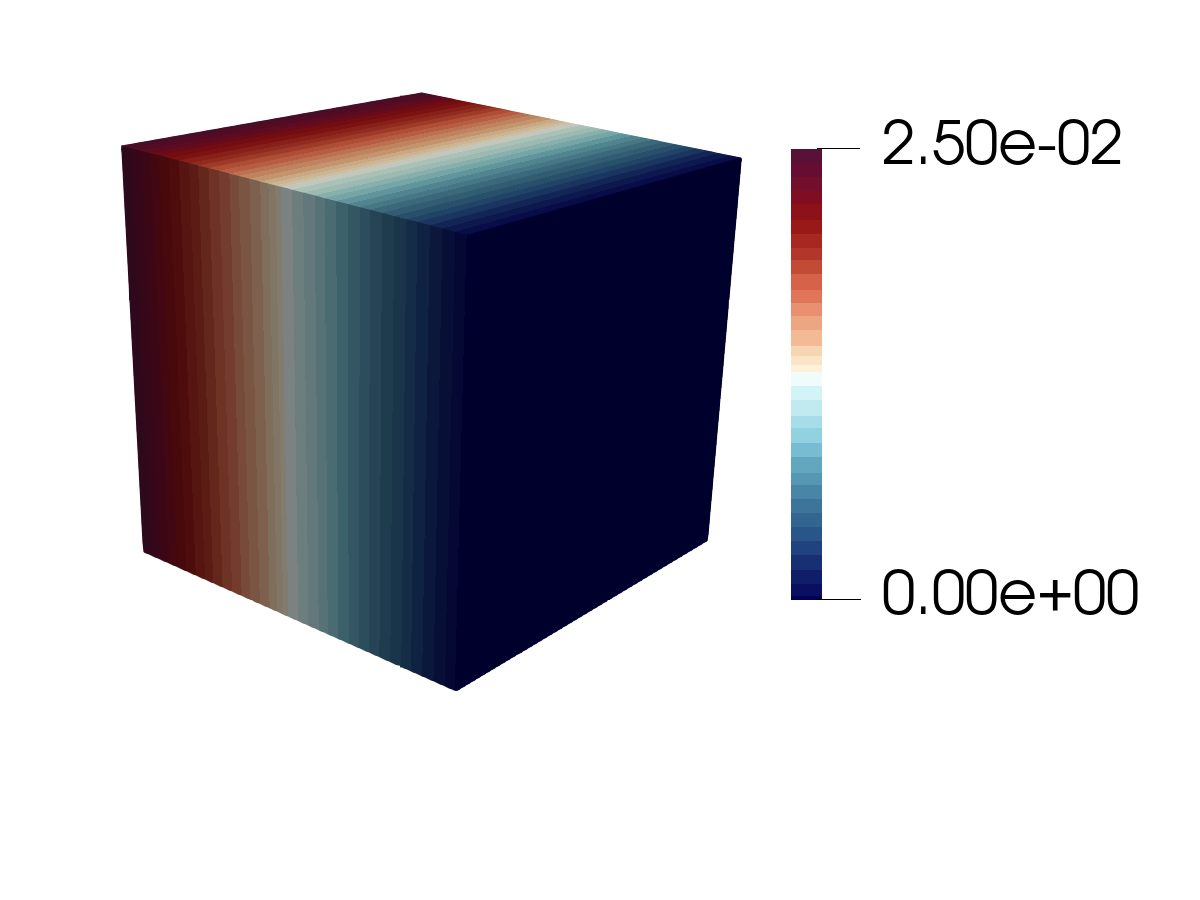}&
        \includegraphics[width=\figsize\linewidth, trim={{\cropL} {\cropB} {\cropR} {\cropT}}, clip]{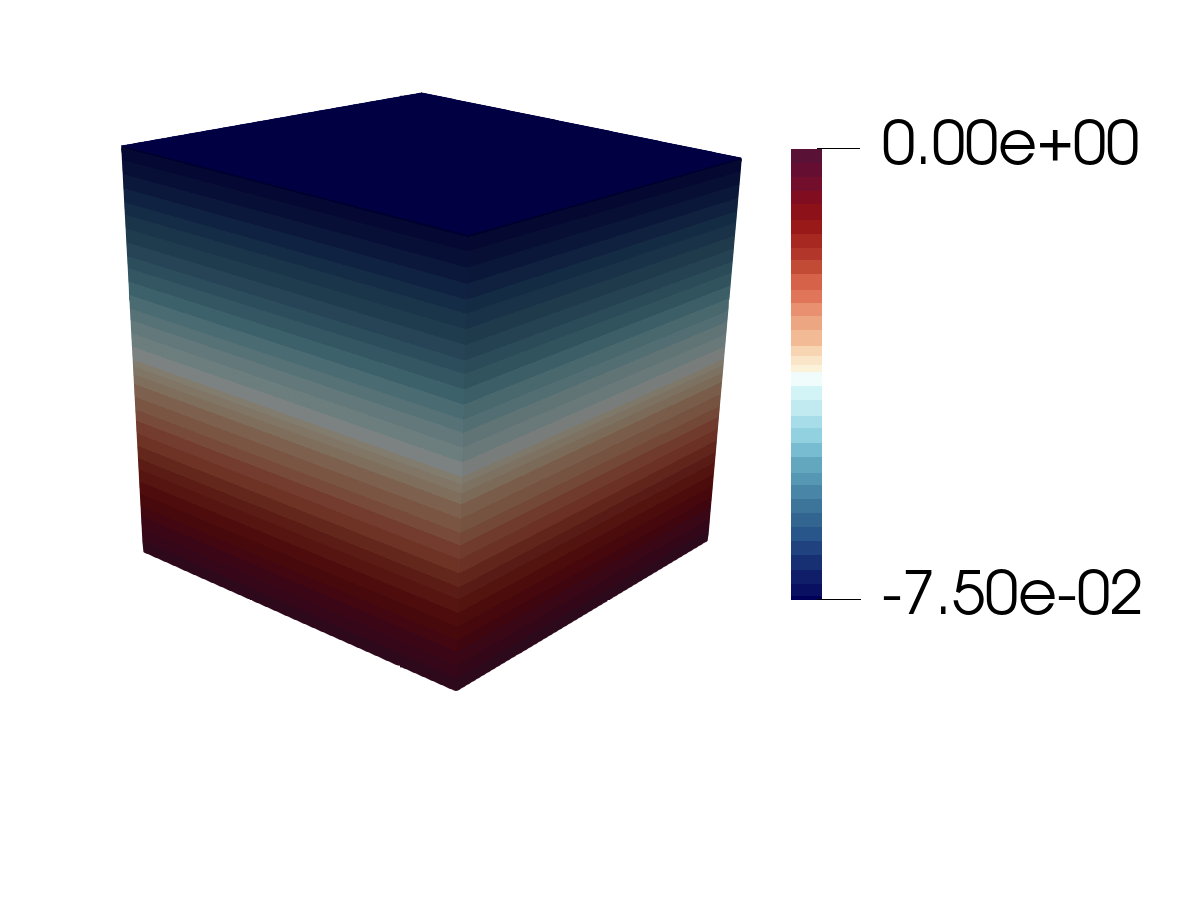}&
        \includegraphics[width=\figsize\linewidth, trim={{\cropL} {\cropB} {\cropR} {\cropT}}, clip]{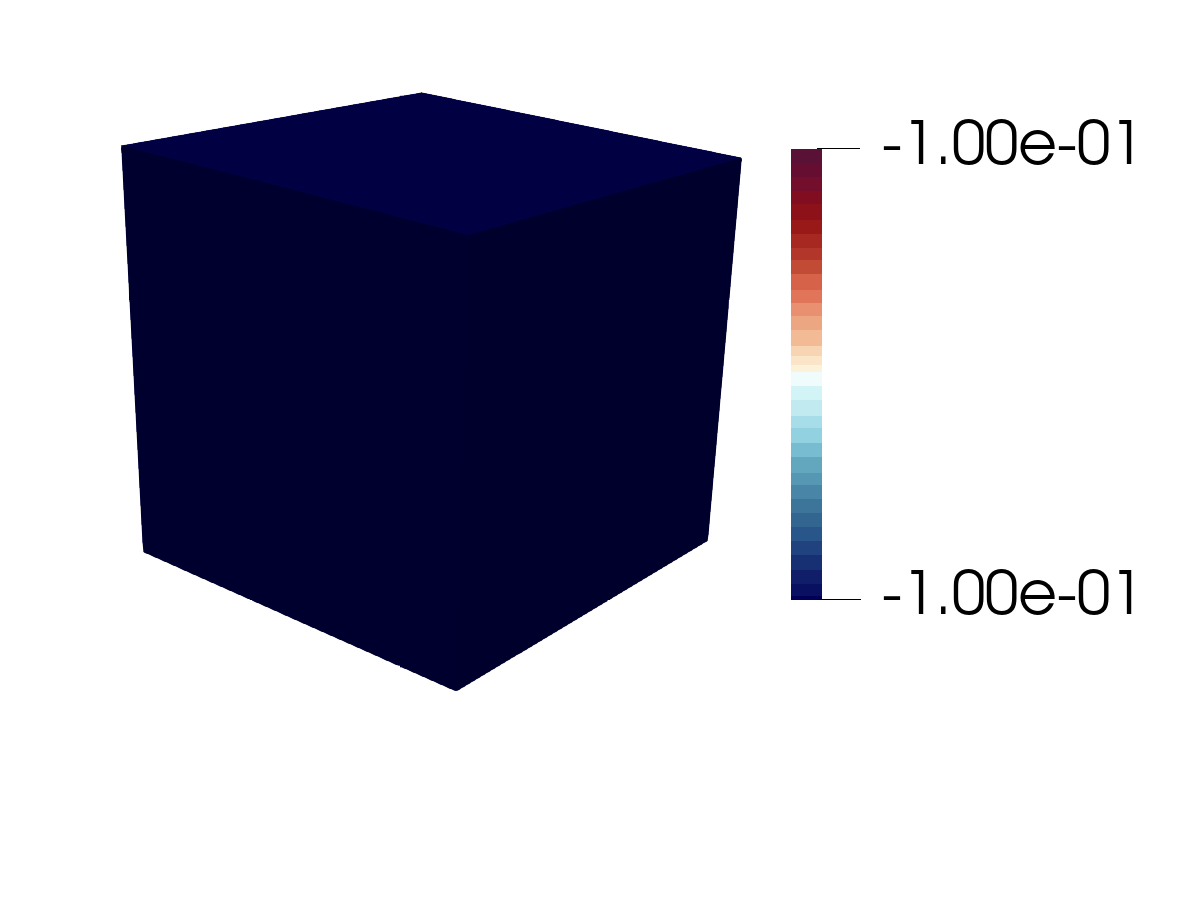}\\

        \rownameform{\hspace{-2.1cm}\scalebox{0.9}{Point-wise error}}&
        \includegraphics[width=\figsize\linewidth, trim={{\cropL} {\cropB} {\cropR} {\cropT}}, clip]{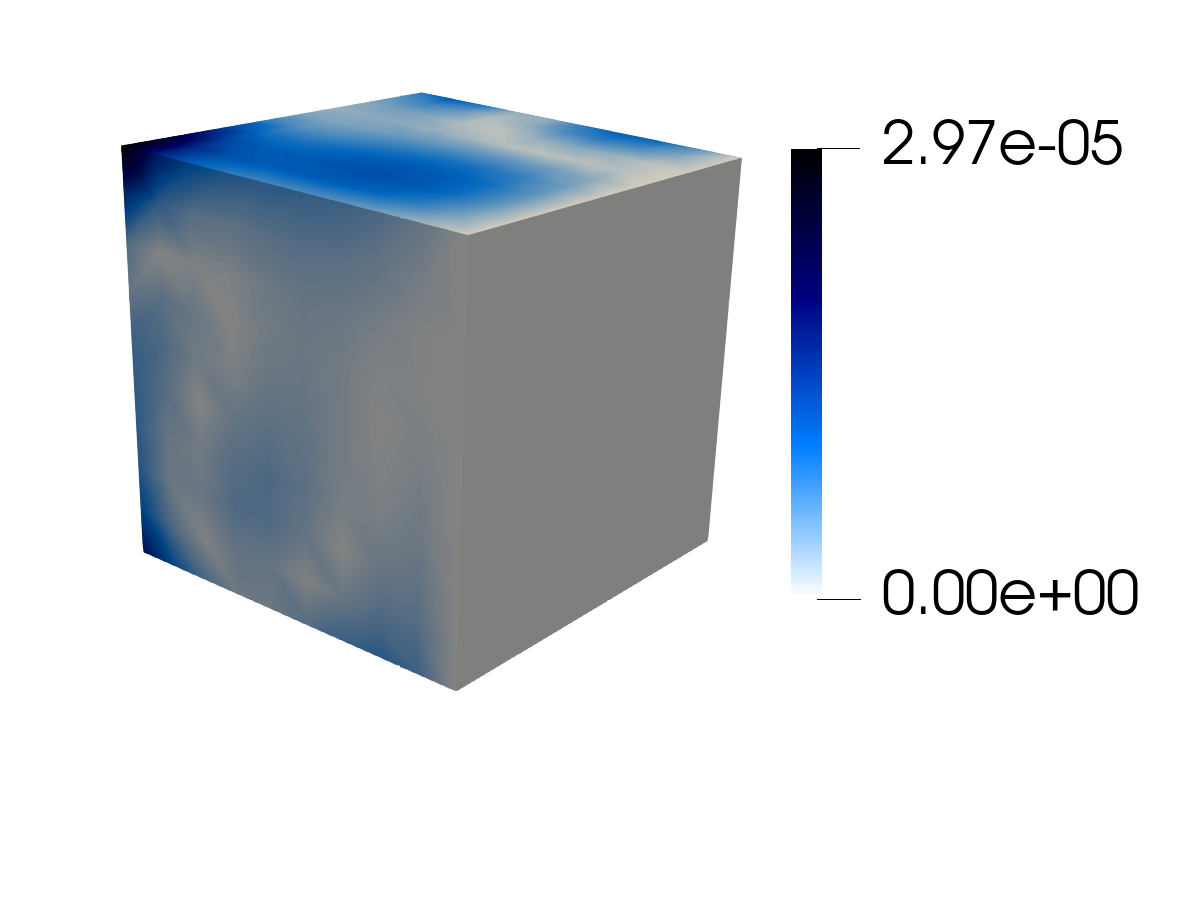}&
        \includegraphics[width=\figsize\linewidth, trim={{\cropL} {\cropB} {\cropR} {\cropT}}, clip]{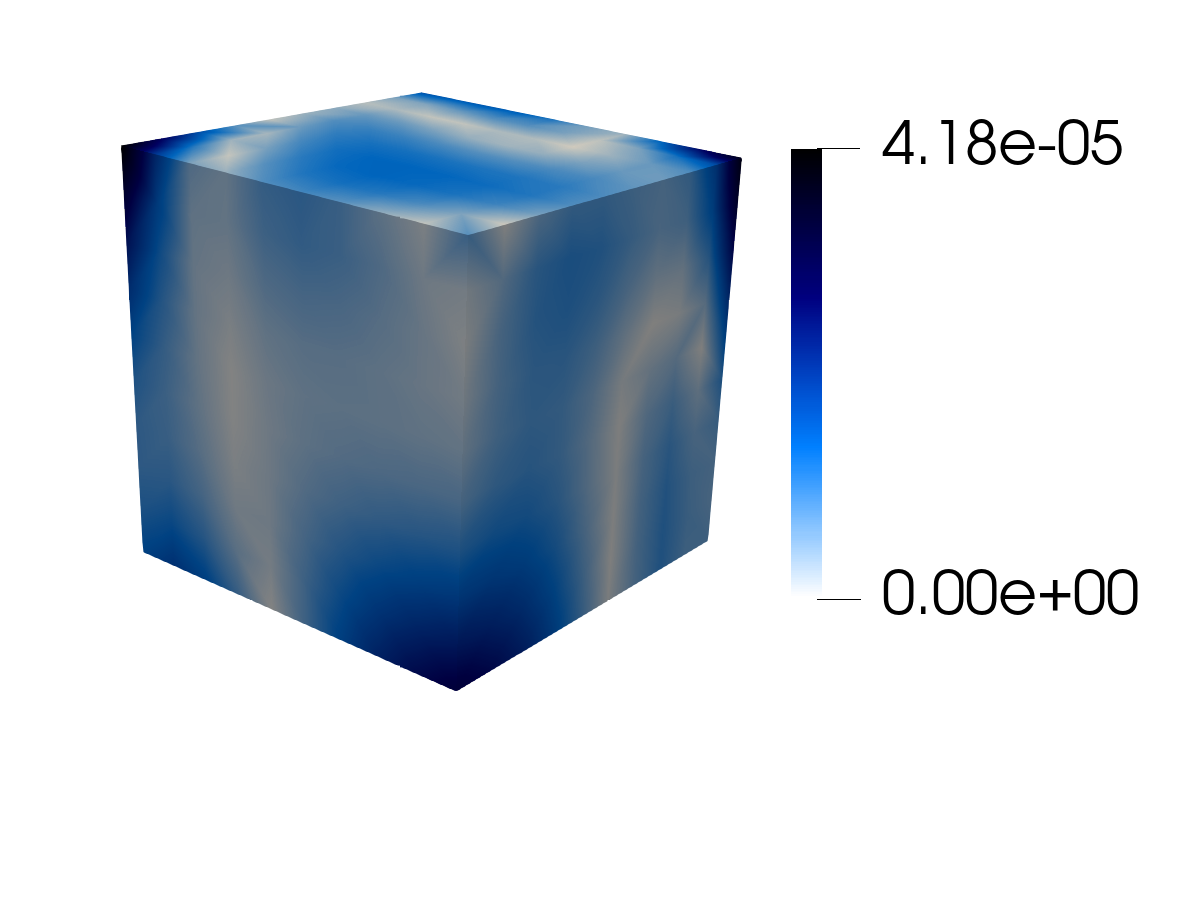}&
        \includegraphics[width=\figsize\linewidth, trim={{\cropL} {\cropB} {\cropR} {\cropT}}, clip]{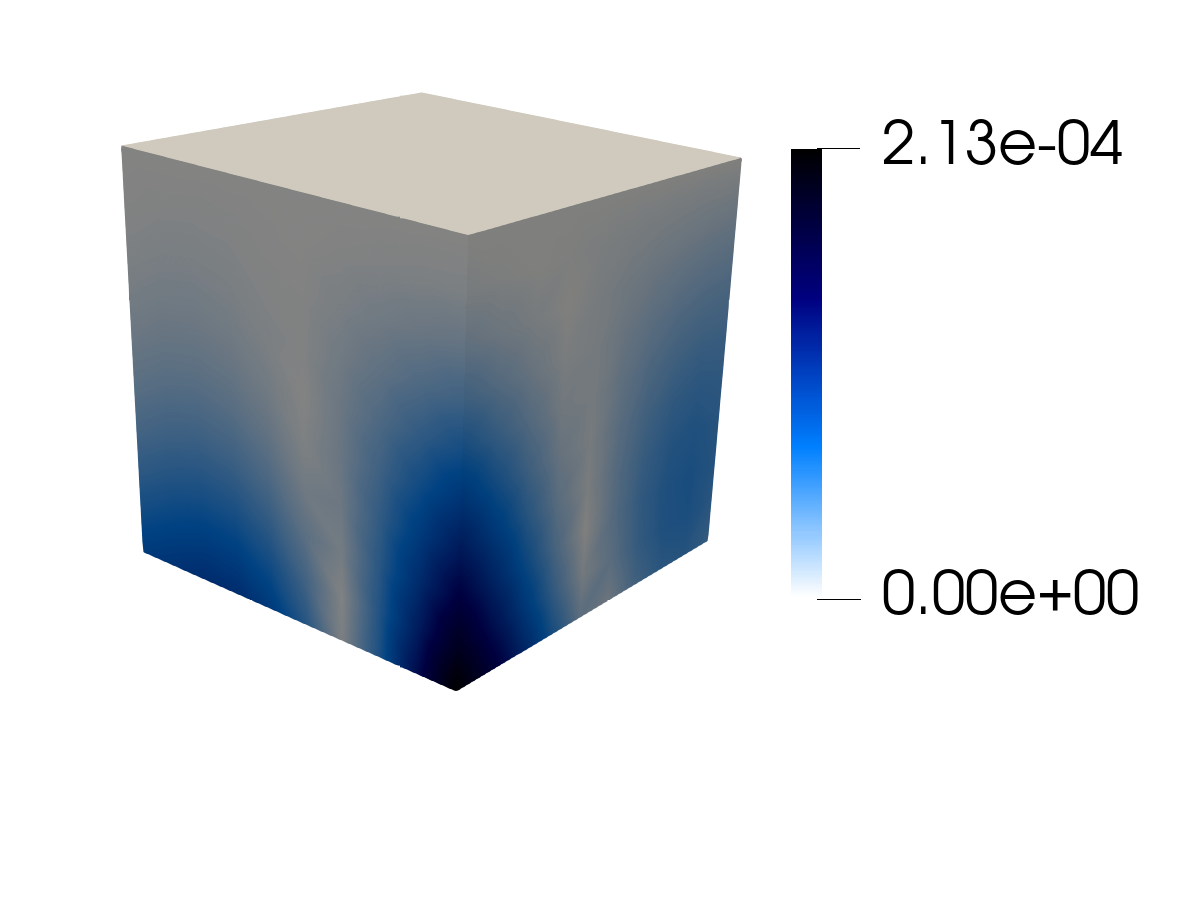}
    \end{tabular}
    \caption{Comparison of the PINN predictions and the analytical solution for the single contact patch test.
    Point-wise error is defined as $E_{abs}^{*}=\text{abs}(\tilde{*} - *)$.}
    \label{fig:result_patch}
\end{figure}
Thanks to the output transformation, the boundary conditions at $x=0$ and $z=0$ are enforced as hard constraints, 
ensuring that the corresponding displacement components are strictly zero. This accuracy is further supported by 
point-wise error analysis, which confirms the close agreement between predictions and the analytical solution. 
Similarly, the normal stress component in the $y$-direction, $\sigma_{yy}$, remains close to the expected value of 
$-0.1$ across the entire domain, consistently aligning with the analytical solution.

\subsection{Hertzian Contact Problem}
\label{sec:Hertzian_contact_problem}
\begin{figure}
    \centering
    \includegraphics[width=0.85\linewidth]{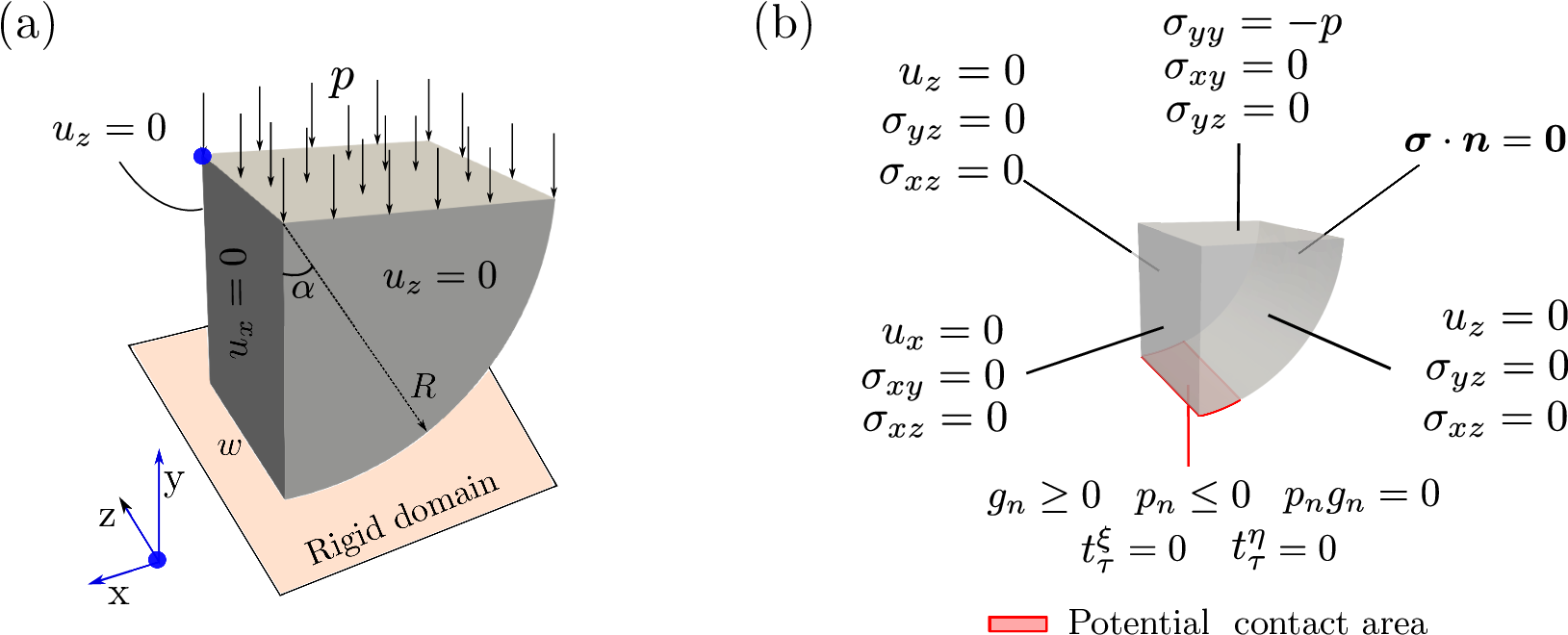}
    \caption{The Hertzian contact problem between an elastic quarter-cylinder and a rigid ﬂat domain:
    (a) geometry, constraints, loading using the symmetry, 
    (b) all relevant boundary conditions.}
    \label{fig:hertzian_contact_3d}
\end{figure}

In the second example, we consider a finite linear elastic half-cylinder ($E=200$ and $\nu=0.3$) lying on a rigid 
flat domain under a uniform pressure $p=0.5$ as depicted in Fig. \ref{fig:hertzian_contact_3d}a.
Additionally, we set the height of the quarter cylinder to $w=1$ and the radius as $R=1$. 
See Appendix~1 for the analytical solution.

Similar to the previous example, the following output transformation is applied to impose the Dirichlet and Neumann boundary 
conditions (BCs) shown in Fig. \ref{fig:hertzian_contact_3d}b as hard constraints:
\begin{align*}
    \pinnUComponent{x} = \frac{x}{E} \pinnUPredicter{x}, \:\ 
    \pinnUComponent{y} = \frac{1}{E} \pinnUPredicter{y}, \:\
    \pinnUComponent{z} = \frac{z(w+z)}{E}& \pinnUPredicter{z}, \:\ 
    \pinnSComponent{yy}= -p + (-y)\pinnSPredicter{yy}, \\ 
    \pinnSComponent{xy}= xy\pinnSPredicter{xy}, \:\
    \pinnSComponent{yz}= (w+z)zy\pinnSPredicter{yz}&, \:\
    \pinnSComponent{xz}= (w+z)zx\pinnSPredicter{xz}.  
\end{align*}
This output transformation ensures that the DBCs and NBCs on the flat surfaces are strictly enforced. 
To simplify the optimization process, the displacements are scaled using the inverse of the Young's modulus.
Moreover, NBCs on the curved surface are enforced using soft constraints due to the coupling between 
the normal and shear stress components on those surfaces. 
To define the potential contact area, we set $\alpha = 15^{\circ}$ and enforce the contact constraints exclusively on the 
points lying on that surface (highlighted in red in Fig~\ref{fig:hertzian_contact_3d}b).  

\begin{figure}
    \def\figsize{0.2}
    \def\hsp{\hspace{-1.cm}}
    \def\hskiplocal{\hskip 0.25cm}
    \centering
    \def\arraystretch{0.2}% 
    \begin{tabular}{cc@{\hskiplocal}cc@{\hskiplocal}cc@{\hskiplocal}cc}
    &\scalebox{1}{Displacement} 
    & 
    &\scalebox{1}{Norm. stress} 
    &
    &\scalebox{1}{Shear stress} 
    &
    &\scalebox{1}{Norm. stress polar}\\[1ex]
    \rownameform{\hspace{-1cm}\scalebox{1}{$x$-direction}}&
    \includegraphics[width=\figsize\linewidth]{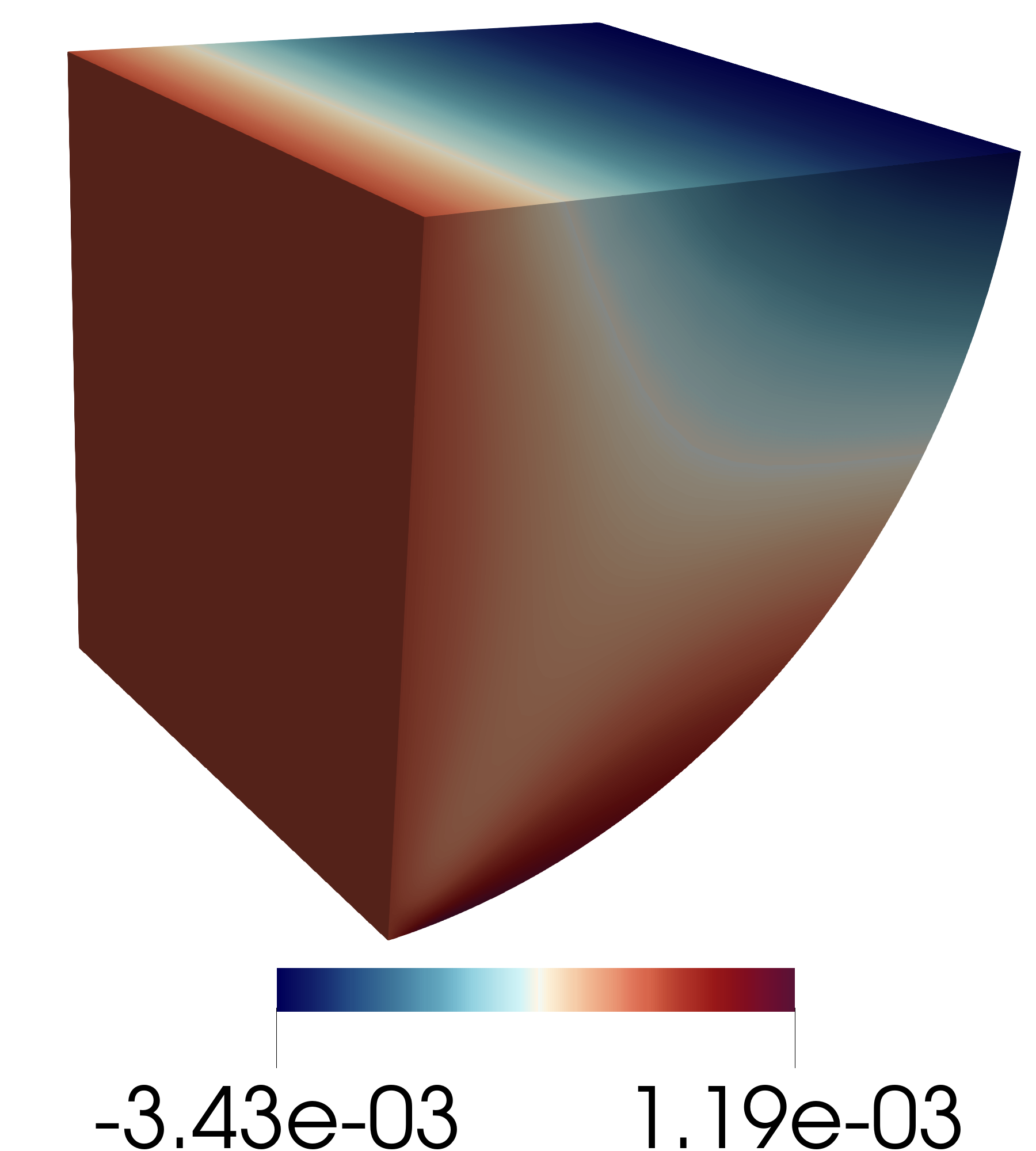}&
    \rownameform{\hspace{-1cm}\scalebox{1}{$xx$-direction}}&
    \includegraphics[width=\figsize\linewidth]{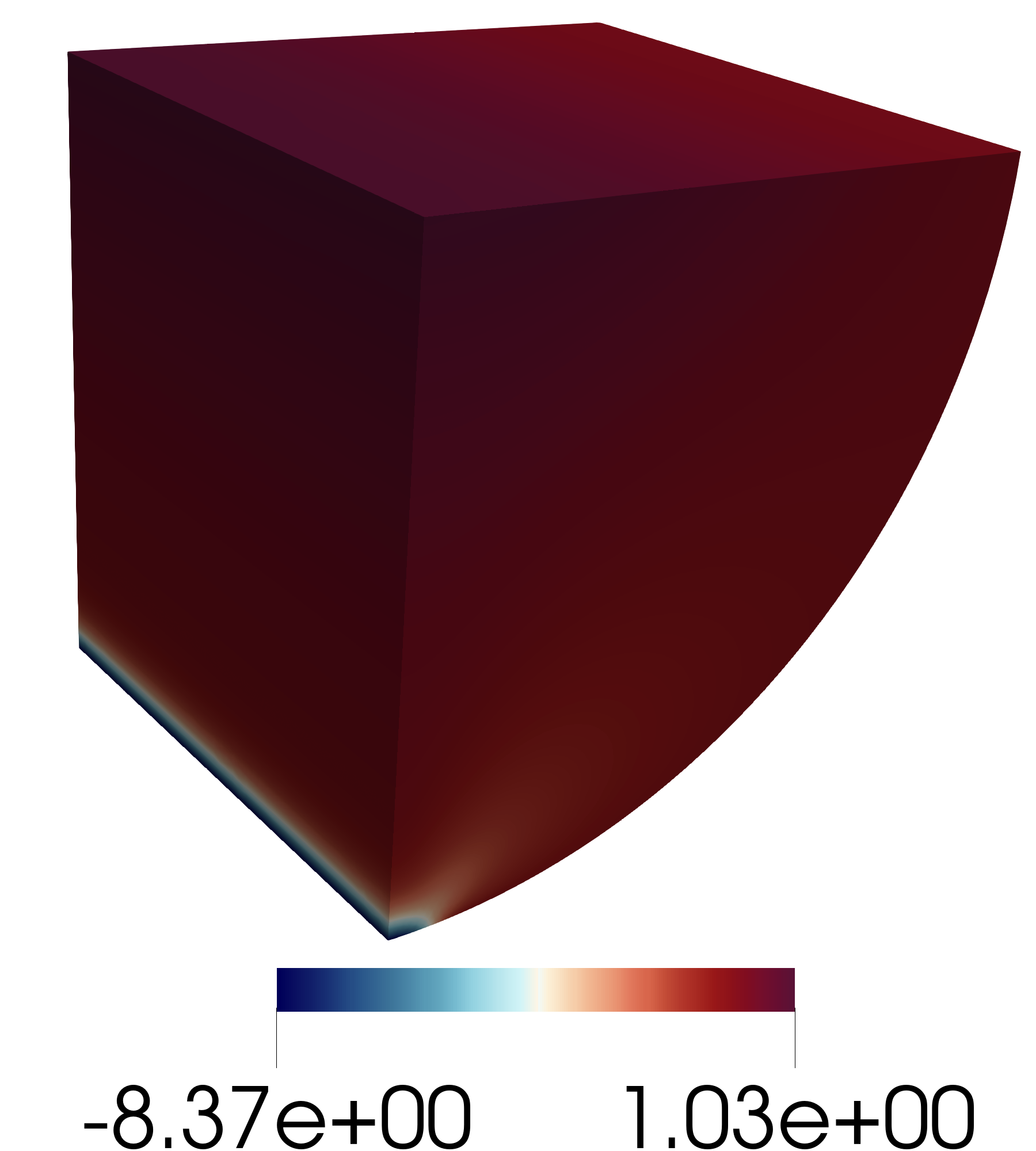}&
    \rownameform{\hspace{-1cm}\scalebox{1}{$xy$-direction}}&
    \includegraphics[width=\figsize\linewidth]{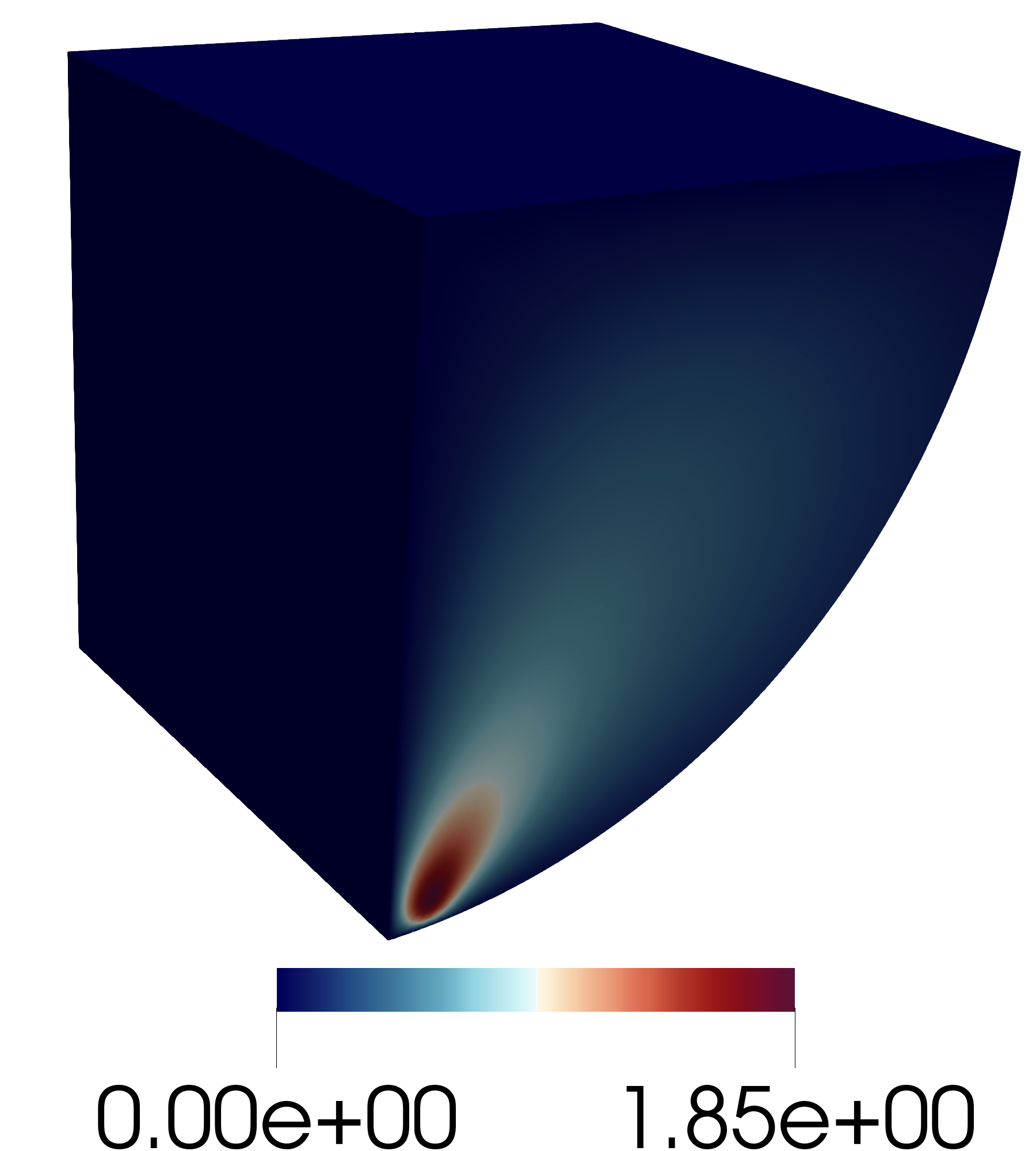}&
    \rownameform{\hspace{-1cm}\scalebox{1}{$rr$-direction}}&
    \includegraphics[width=\figsize\linewidth]{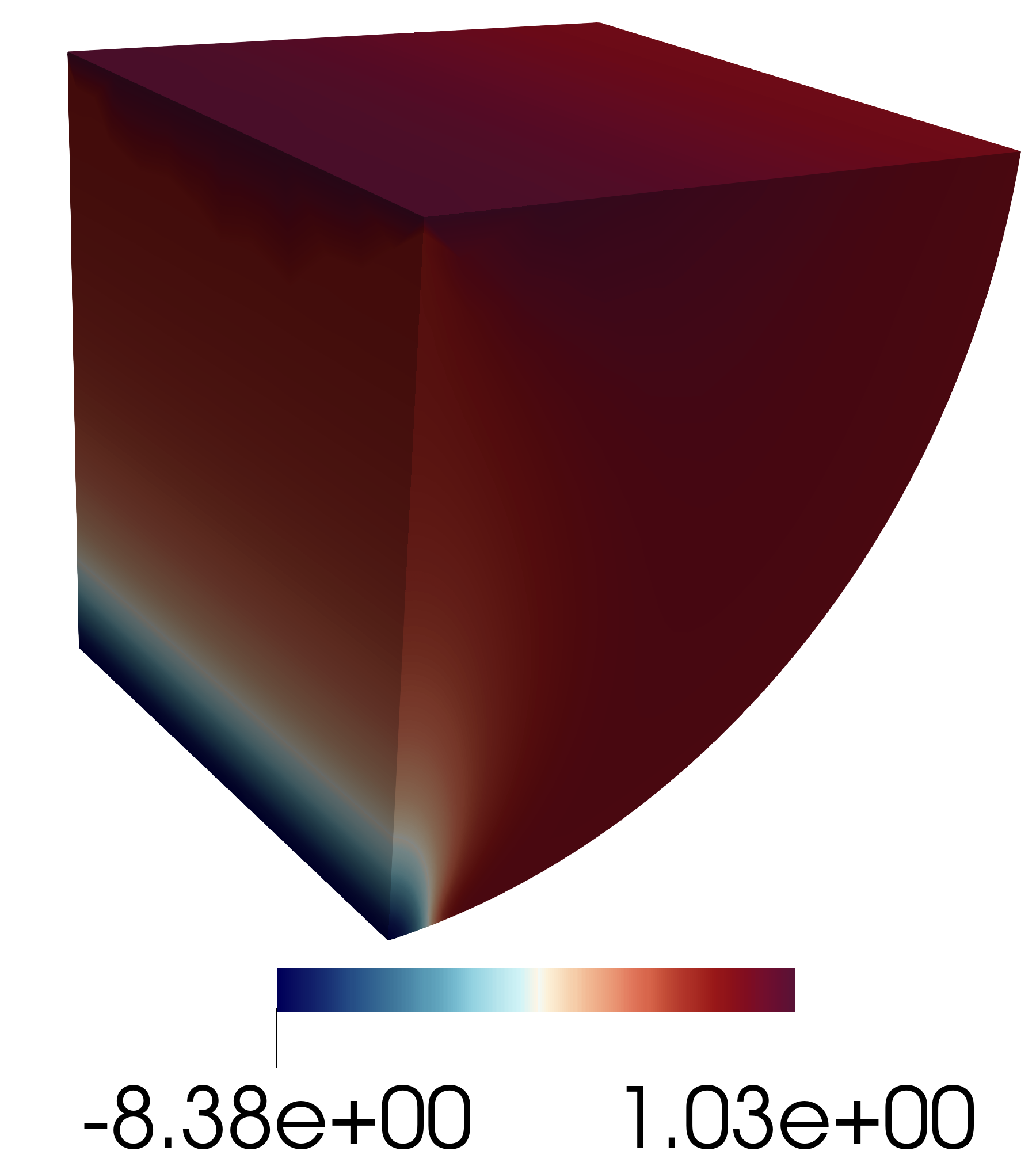}\\[1ex]
    \rownameform{\hspace{-1cm}\scalebox{1}{$y$-direction}}&
    \includegraphics[width=\figsize\linewidth]{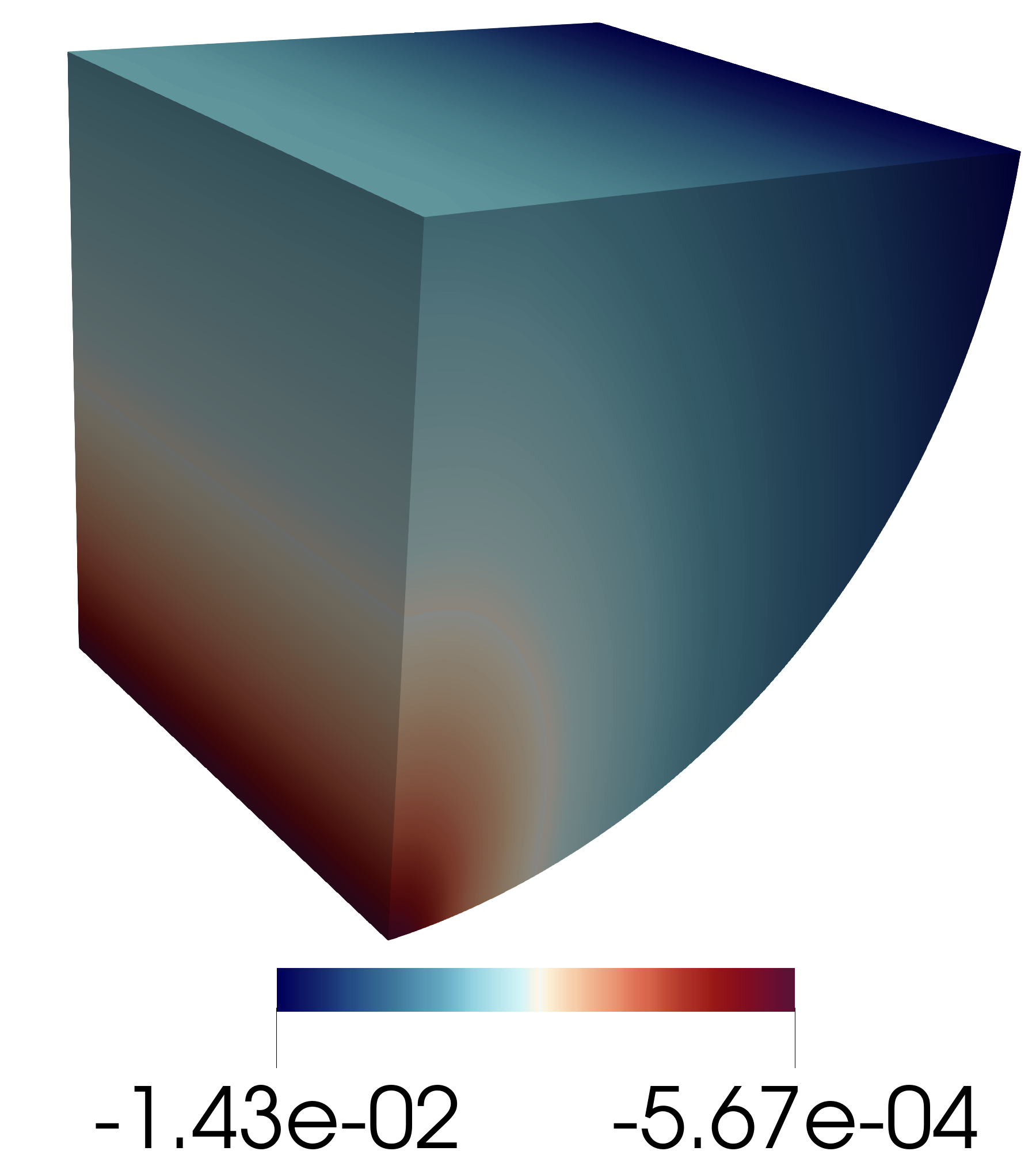}&
    \rownameform{\hspace{-1cm}\scalebox{1}{$yy$-direction}}&
    \includegraphics[width=\figsize\linewidth]{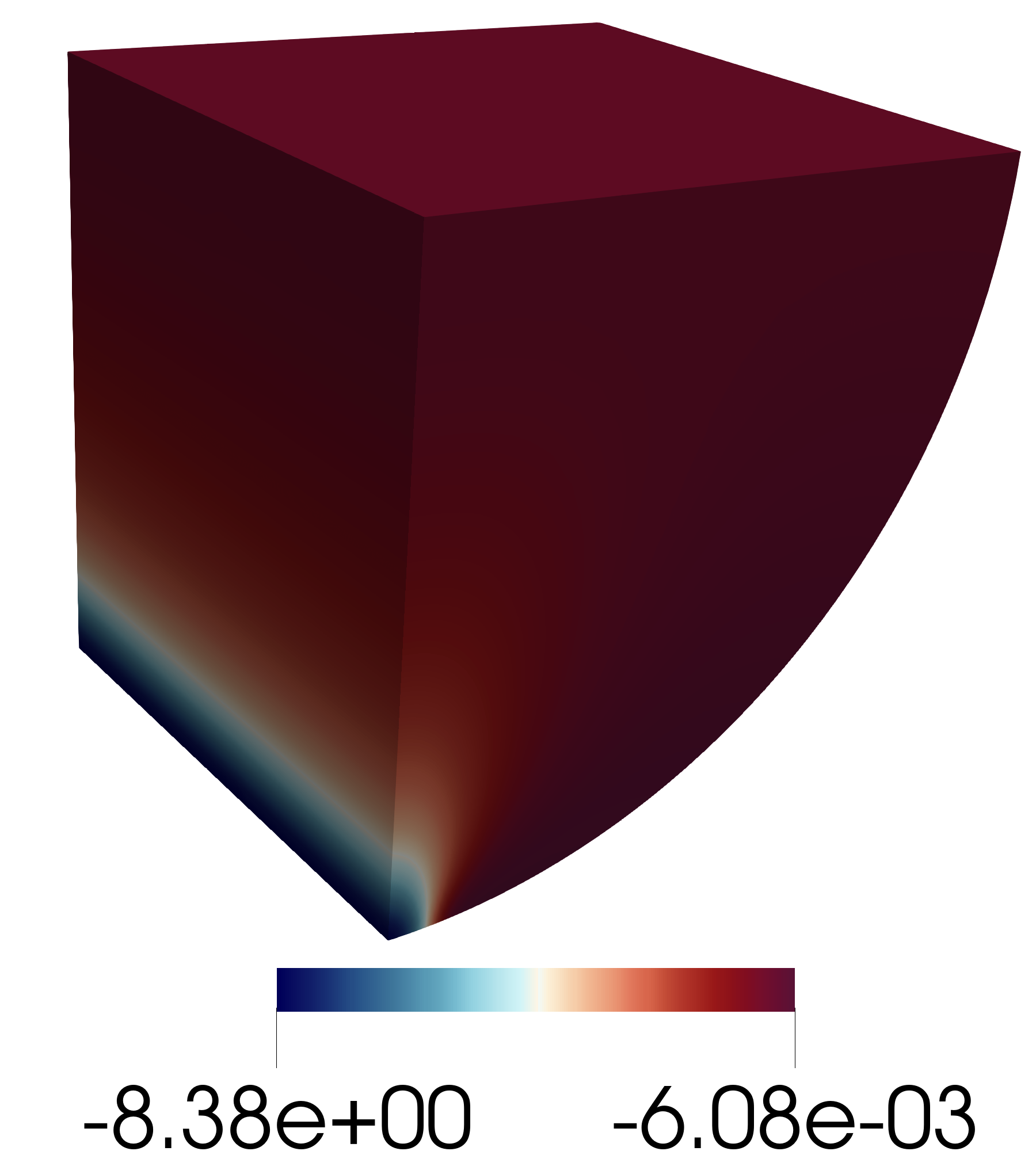}&
    \rownameform{\hspace{-1cm}\scalebox{1}{$yz$-direction}}&
    \includegraphics[width=\figsize\linewidth]{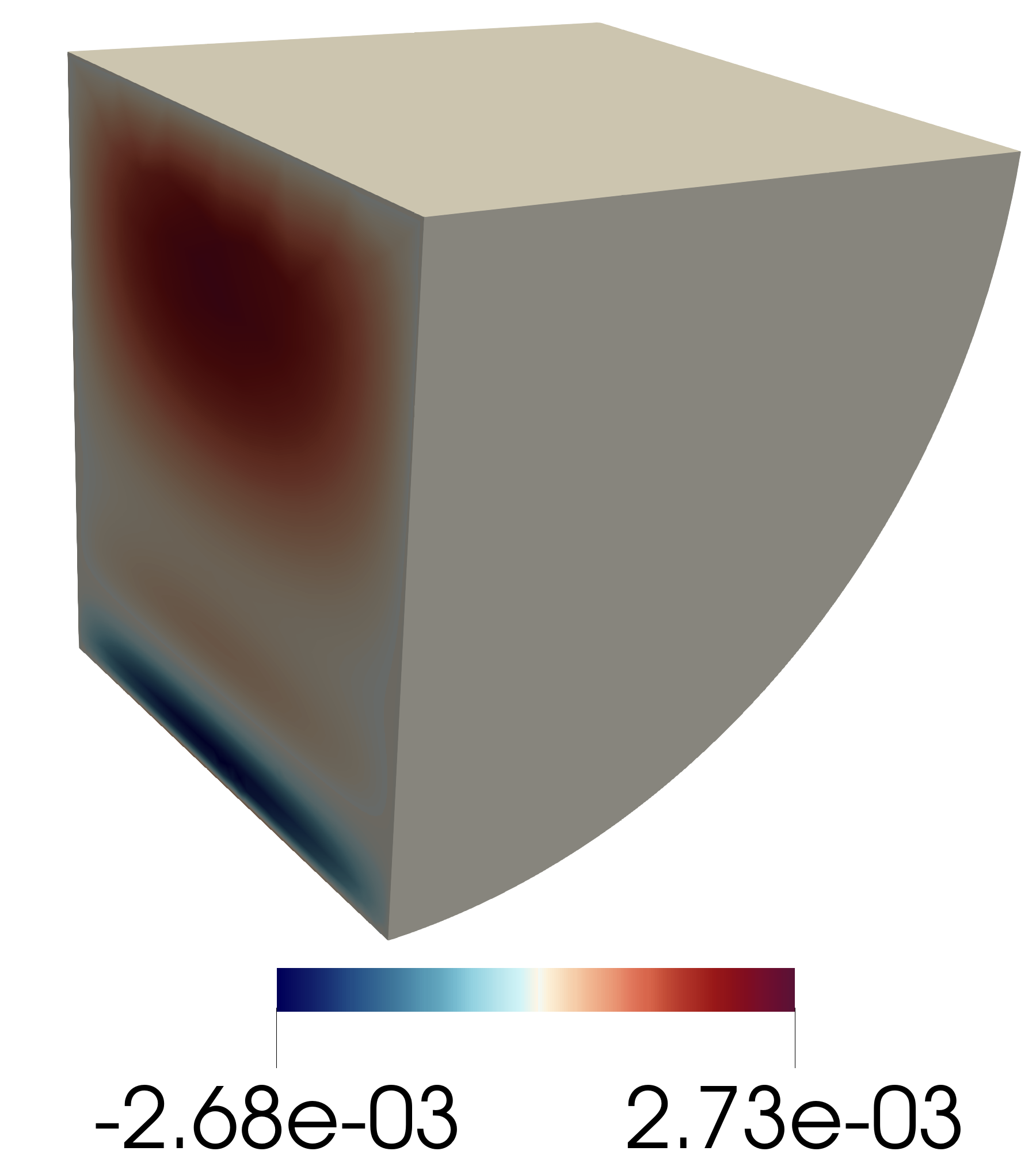}&
    \rownameform{\hspace{-1cm}\scalebox{1}{$\theta \theta$-direction}}&
    \includegraphics[width=\figsize\linewidth]{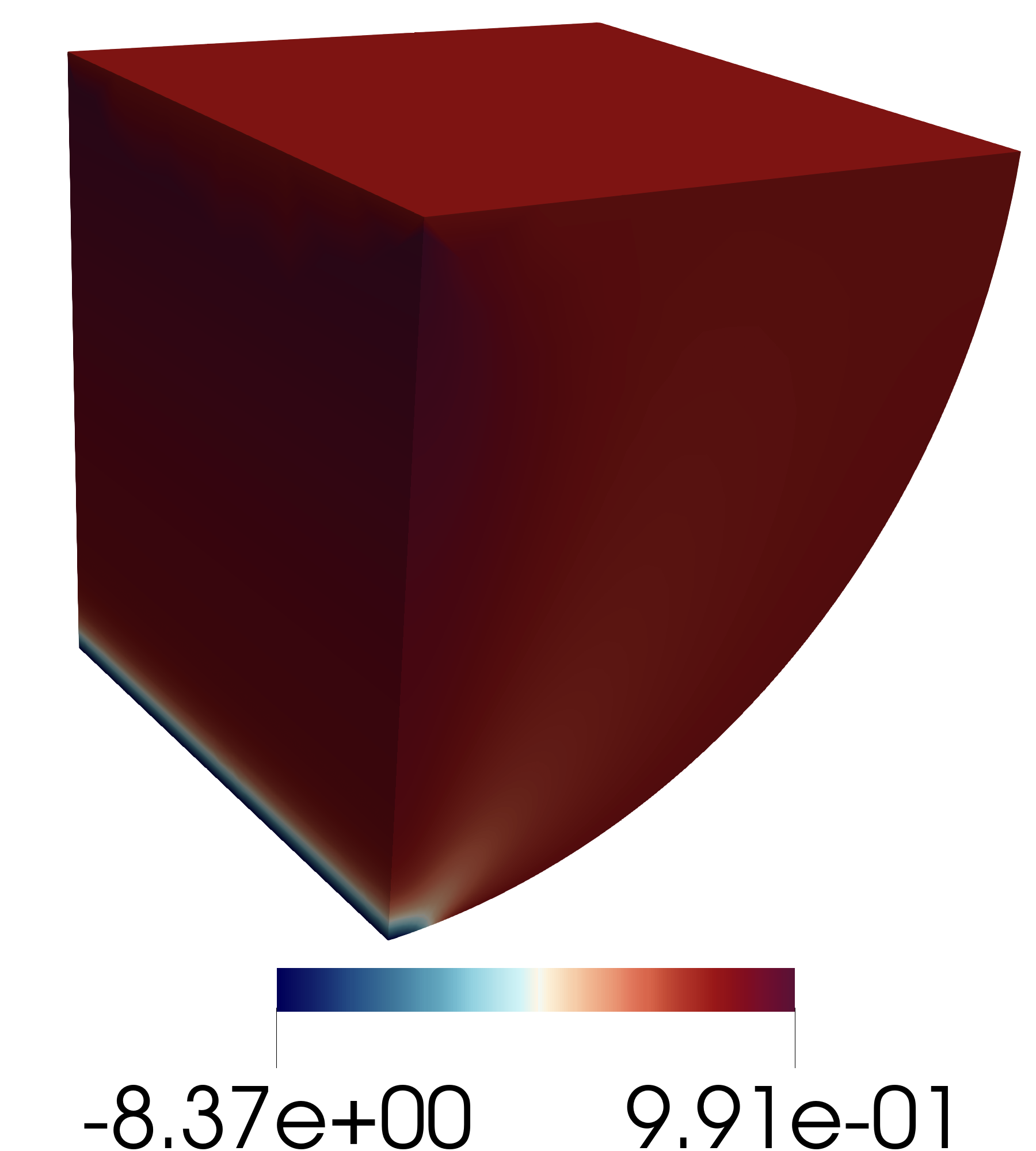}\\[1ex]
    \rownameform{\hspace{-1cm}\scalebox{1}{$z$-direction}}&
    \includegraphics[width=\figsize\linewidth]{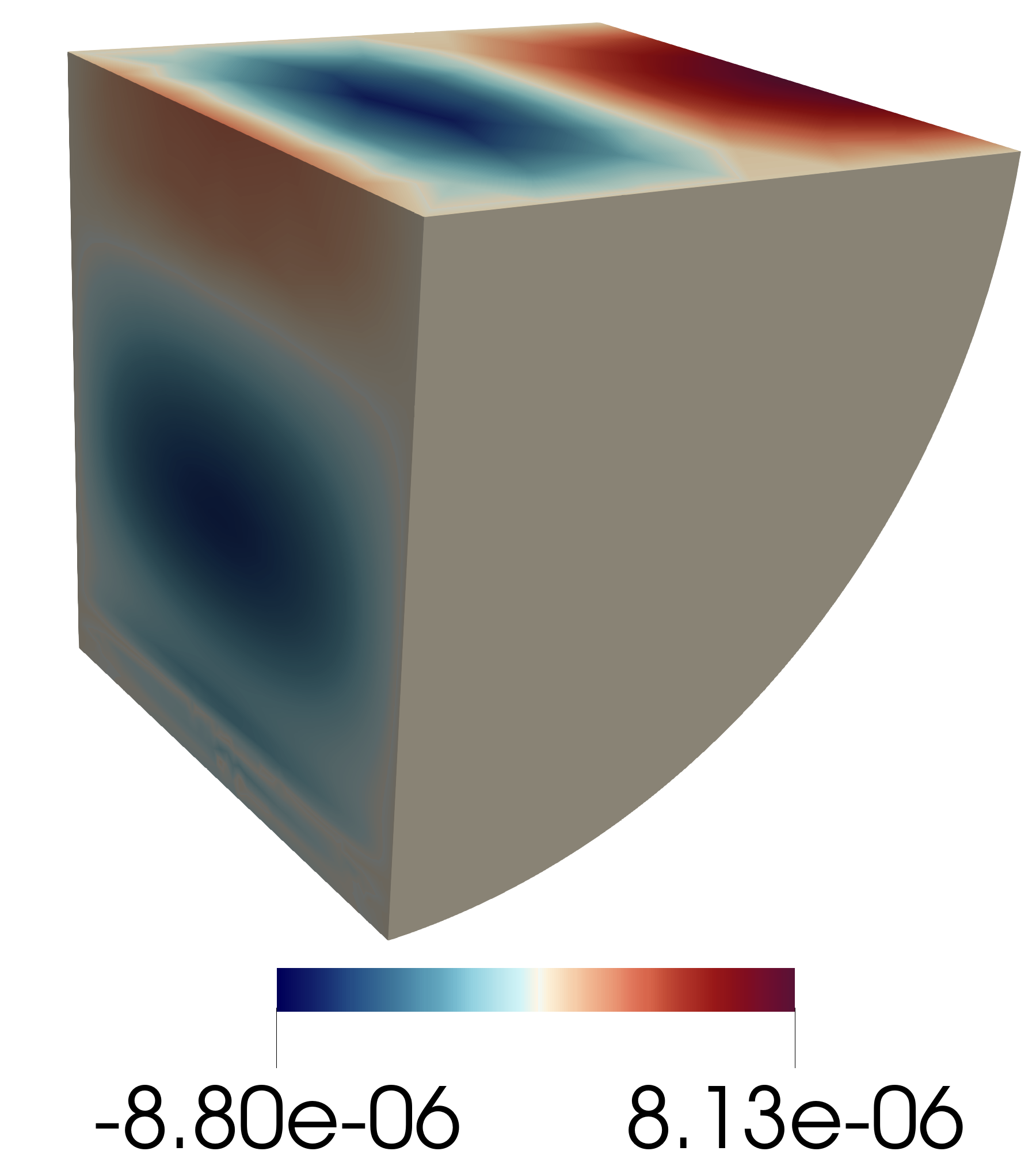}&
    \rownameform{\hspace{-1cm}\scalebox{1}{$zz$-direction}}&
    \includegraphics[width=\figsize\linewidth]{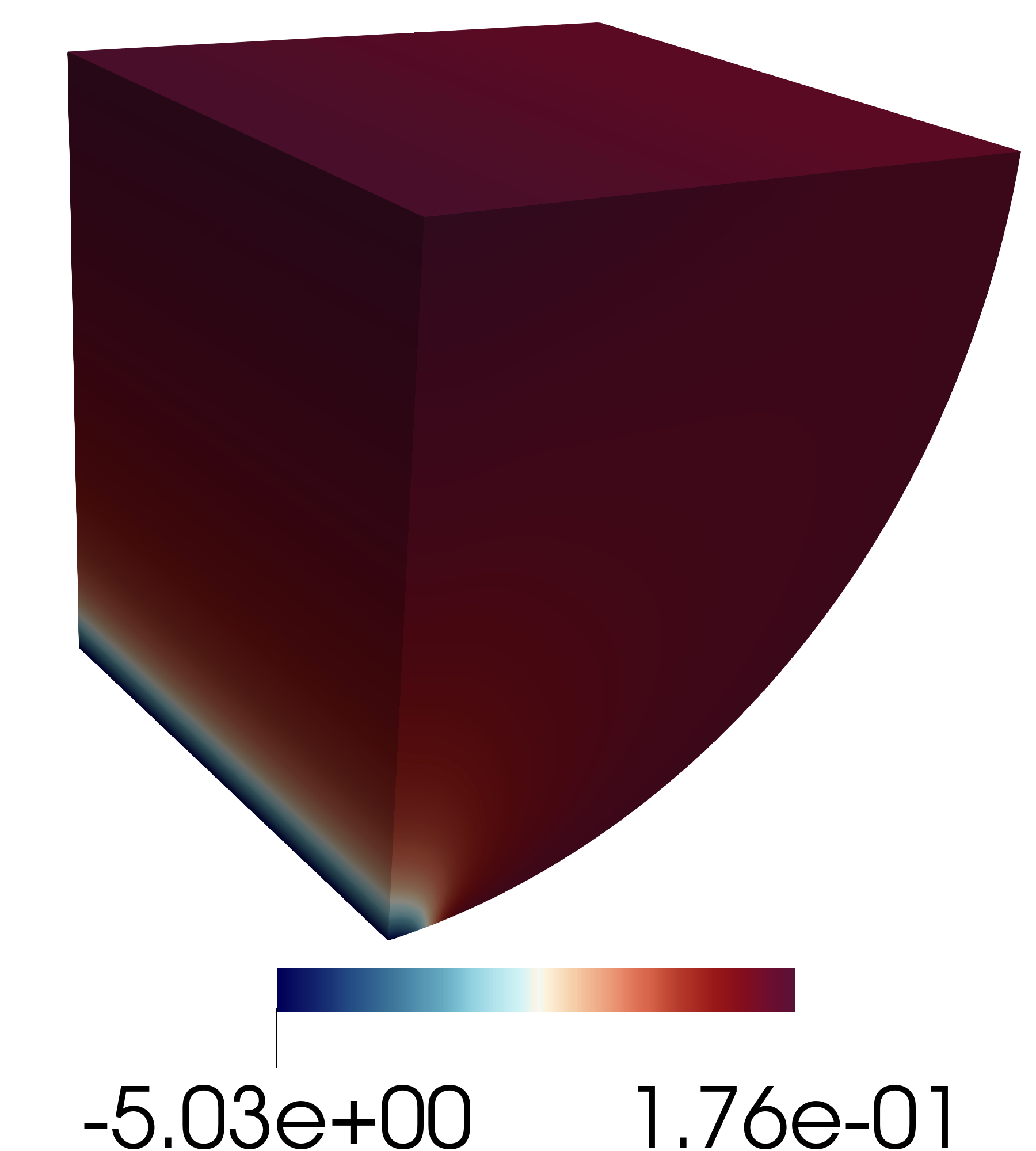}&
    \rownameform{\hspace{-1cm}\scalebox{1}{$xz$-direction}}&
    \includegraphics[width=\figsize\linewidth]{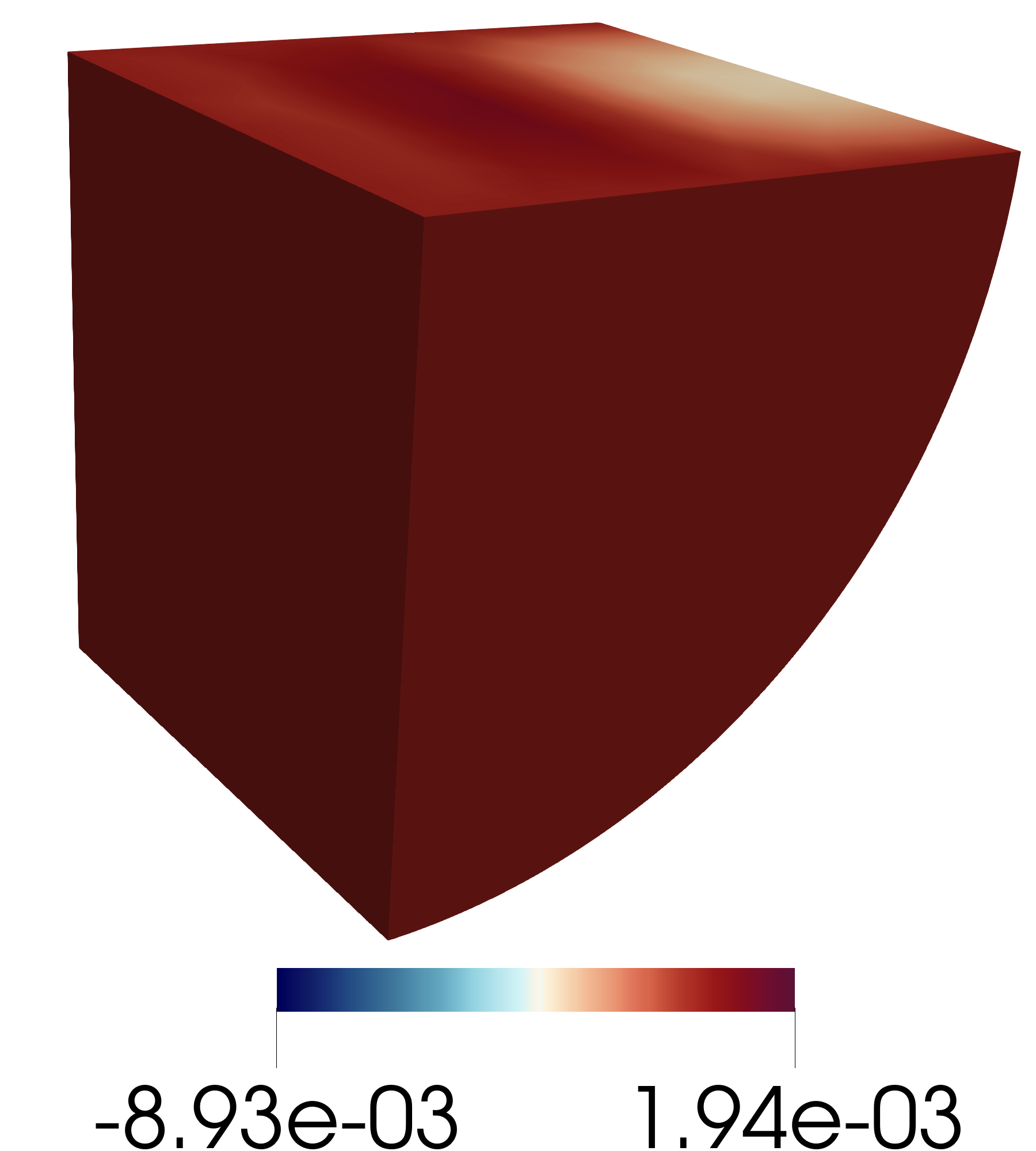}&
    \rownameform{\hspace{-1cm}\scalebox{1}{$zz$-direction}}&
    \includegraphics[width=\figsize\linewidth]{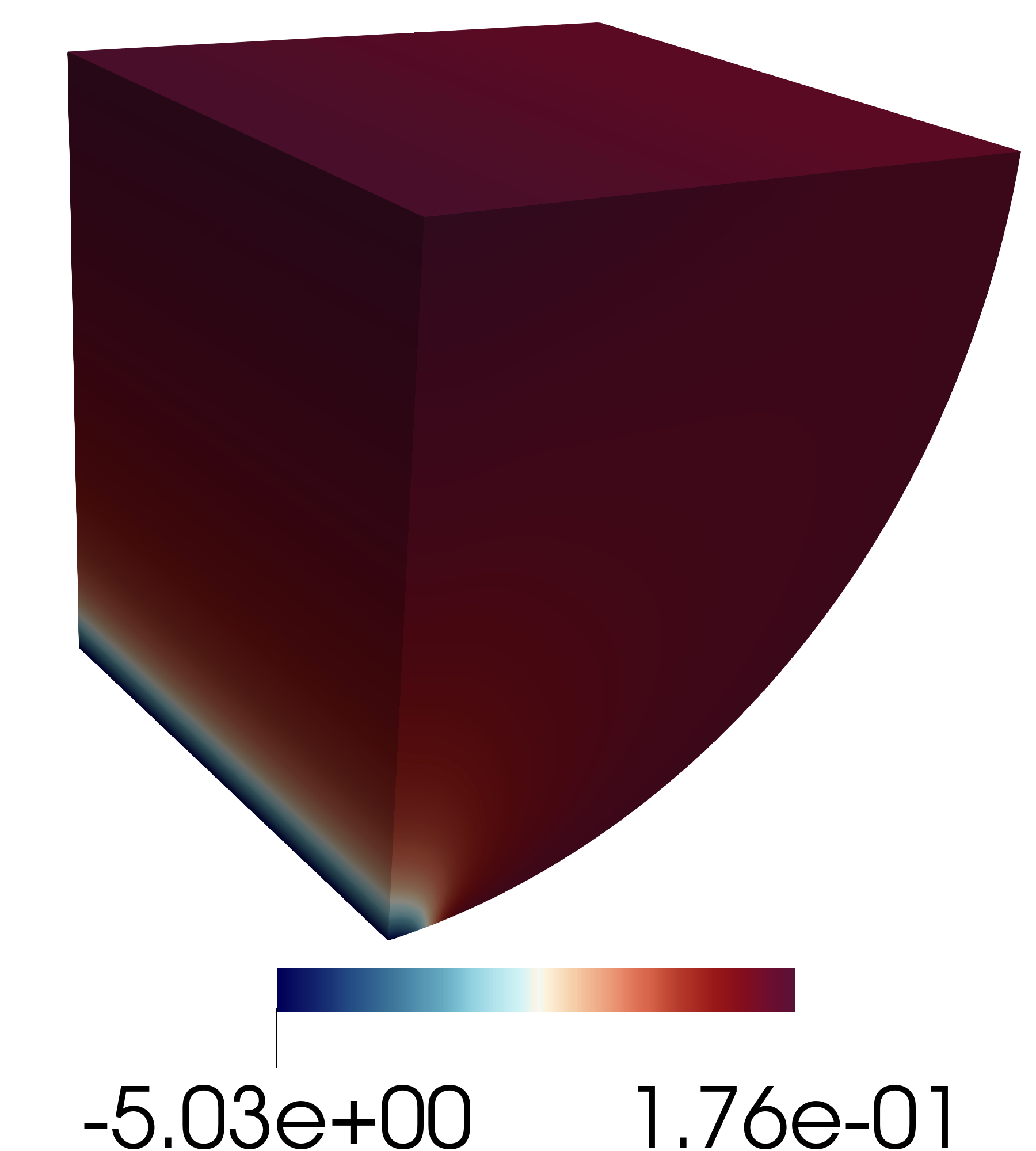}\\%[-1ex]
\end{tabular}
    \caption{The data-enhanced PINN predictions for the Hertzian contact example. Polar components are computed based on the cylindrical coordinates.}
    \label{fig:result_hertzian}
\end{figure}

In the following, we investigate two distinct PINN application cases for the Hertzian contact problem. In the first case, we deploy a PINN as 
pure PDE solver (denoted as "plain vanilla PINN"), and in the second case, we deploy a data-enhanced PINN including additional data obtained from the analytical solution 
(see Eqs.~\ref{eq:s_xx}-\ref{eq:s_zz} provided in Appendix~1). In total 150 additional data points are selected along three distinct lines (each involves 50 points)
located at $z=-1$, $z=-0.5$, $z=0$. Each line starts at $y=-1$ and ends at $y=-0.7642$. 
Furthermore, the loss weight for the KKT term in Eq.~\ref{eq:fischer_burmeister_loss} is set 
to $w_{\mathrm{KKT}} = 500$, ensuring that the PINN model prioritizes 
learning the contact constraints.
The same neural network architecture as in the previous example is utilized. 

The PINN predictions for the Hertzian contact example with data enhancement are provided in Fig.~\ref{fig:result_hertzian}. These results 
include the displacement fields and stress fields in both normal and shear directions. Additionally, the normal components of the stress 
tensor are presented in cylindrical coordinates. It can be observed that the employed output transformation explicitly enforces components 
of the displacement and stress fields on the target surfaces, such as $u_z=0$ and $\sigma_{yy}=-0.5$. Furthermore, the maximum pressure 
reaches $\sigma_{rr}=8.38$, which is very close to the maximum analytical pressure $p_{\max}=8.36$ (Eq.~\ref{eq:max_pressure}).
Additionally, see Appendix~3 for the predictions of the plain vanilla PINN.

\begin{figure}
    \centering
    \includegraphics[width=0.95\linewidth]{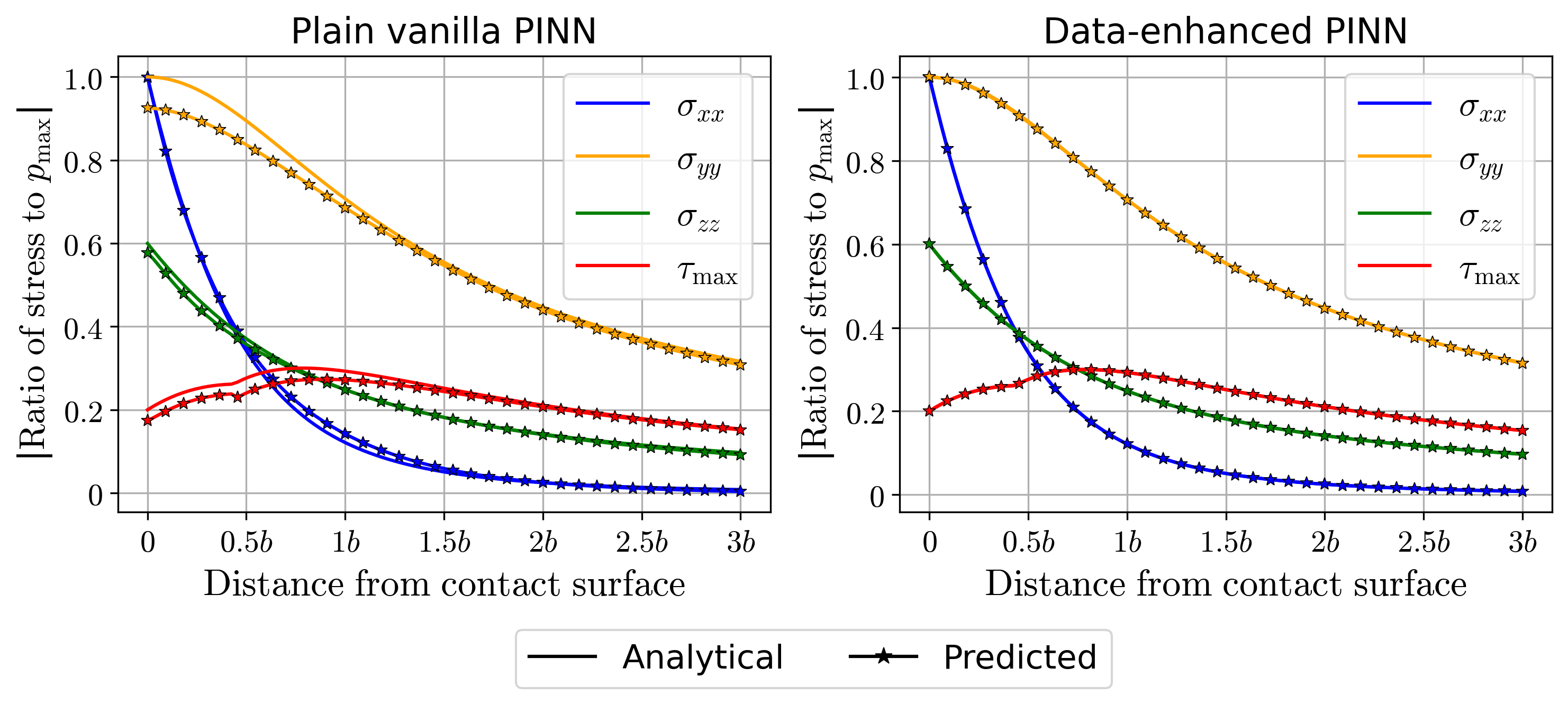}
    \caption{Comparison of the predictions with the analytical solution for the Hertzian contact problem using the plain vanilla PINN and data-enhanced PINN.}
    \label{fig:hertzian_contact_comparison}
\end{figure}

A comparison between the PINN predictions and the analytical solution for both the plain vanilla and data-enhanced PINNs is presented 
in Fig.\ref{fig:hertzian_contact_comparison}. While the plain vanilla PINN underpredicts the analytical solution particularly in regions with higher nonlinearity, 
the data-enhanced PINN demonstrates superior performance, aligning closely with the analytical solution. This can be verified by 
the relative $L_2$ errors as provided in Table \ref{tab:comparison_pinn_analytical} (see Appendix~2). For instance, $\relErrorSComponent{yy}$ is computed as 5.29\%
for the plain vanilla PINN, whereas it is significantly reduced to 0.17\% for the data-enhanced PINN, highlighting the impact of data enhancement.  
Note that the predictions are evaluated using unseen data points at the line $z=-0.75$ to demonstrate the model's ability to generalize. 

% \subsubsection{Case 2: Inverse solver for parameter identification}
% \label{sec:case_2}

\section{Conclusion}
In this study, we have introduced an extension of physics-informed neural networks (PINNs) 
for solving 3D forward problems in contact mechanics under the assumption of linear elasticity. 
The proposed framework has been evaluated on two benchmark examples: the single contact patch test 
and the Hertzian contact problem. Results were validated using existing analytical solutions. 
To enforce the inequality constraints inherent to contact problems, a nonlinear complementarity 
problem (NCP) function, specifically the Fischer-Burmeister function, was employed. 
We conclude that enhancing PINNs with additional measurement data can 
significantly improve the training performance. This leads to hybrid models that effectively leverage both 
the underlying physics and data-driven insights, resulting in improved accuracy and robustness.

\begin{acknowledgement}
% This research paper is funded by dtec.bw - Digitalization and Technology Research Center of the 
% Bundeswehr. dtec.bw is funded by the European Union - NextGenerationEU.
This research paper is funded by dtec.bw - Digitalization and Technology Research Center of the 
Bundeswehr under the project RISK.twin. dtec.bw is funded by the European Union—NextGenerationEU.%\newline
\end{acknowledgement}
\ethics{Competing Interests}{
% This research paper is funded by dtec.bw - Digitalization and Technology Research Center of the 
% Bundeswehr under the project RISK.twin. dtec.bw is funded by the European Union—NextGenerationEU.\newline
The authors have no conflicts of interest to declare that are relevant to the content of this chapter.}

\eject

% \ethics{Ethics Approval}{If your chapter includes primary studies with humans please declare the adherence of ethical standards. Example text: This study was performed in line with the principles of the Declaration of Helsinki. Approval was granted by the Ethics Committee of University B (Date.../No. ...).\newline 
%  In addition, for human participants, authors are required to include a statement that informed consent (to participate and/or to publish) was obtained from individual participants or parents/guardians if the participant is minor or incapable.\newline
% If animals are studied, authors should make sure that the legal requirements or guidelines in the country and/or state or province for the care and use of animals have been followed or specify that no ethics approval was required.}
%%%%%%%%%%%%%%%%%%%%%%%%%%%%%%%%%%%%%%%%%%%%%%%%%%%%%%%%%%%%%%%%%%%%%%%%%%%%%%%%%%%%%%%%%%%%%%%%%%%%%%%%%%%%%%%%%%%%%%%%%%%%%%%%%%%%%%%%%%%%%%%%%%%%
\section*{Appendix 1}%\label{sec:appendix}
%\addcontentsline{toc}{section}{Appendix}
The analytical solution for the Hertzian contact problem between a cylinder and a plane rigid surface is expressed as \cite{budynas2014} 
\begin{align}
    \label{eq:s_xx}
    \sigma_{xx} & = -p_{\max }\left(\frac{1+2 \dfrac{y^2}{b^2}}{\sqrt{1+\dfrac{y^2}{b^2}}}-2\left|\frac{y}{b}\right|\right), \\
    \label{eq:s_yy}
    \sigma_{yy} & = \frac{-p_{\max}}{\sqrt{1+\dfrac{y^2}{b^2}}}, \\
    \label{eq:s_zz}
    \sigma_{zz} & = -2 \nu p_{\max }\left(\sqrt{1+\frac{y^2}{b^2}}-\left|\frac{y}{b}\right|\right),
\end{align}
where 
\begin{equation}
    \label{eq:max_pressure}
    p_{\max} = \frac{2F}{\pi b w}, \quad 
    b = \sqrt{\frac{2F}{\pi w} \frac{(1-\nu^2)/E}{1/(2R)}}, \quad
    F = 2Rwp.
    %\:\ 
\end{equation}
% \begin{align}
%     \label{eq:max_pressure}
%     p_{\max} & = \frac{2F}{\pi b w}, \\
%     b & = \sqrt{\frac{2F}{\pi w} \frac{(1-\nu^2)/E}{1/(2R)}},\\
%     F &= 2Rwp.
% \end{align}
By inserting $p=0.5$, $R=1$, $w=1$, $\nu=0.3$ and $E=200$, the area of contact, which is a narrow rectangle, i.e. $bw$, can be calculated,
leading to $b=0.076$, and the maximum pressure is computed as $p_{\max}=8.36$. Additionally, 
the maximum shear stress can be expressed as 
\begin{align}
    \tau_{\max} &= \frac{\sigma_{zz} - \sigma_{yy}}{2} \quad \text{for} \quad 0 \leq y \leq 0.436b, \\ 
    \tau_{\max} &= \frac{\sigma_{xx} - \sigma_{yy}}{2} \quad \text{for} \quad y \geq 0.436b.
\end{align}
The analytical solutions can be easily computed based on the vertical distance from the contact zone $y$. 
%%%%%%%%%%%%%%%%%%%%%%%%%%%%%%%%%%%%%%%%%%%%%%%%%%%%%%%%%%%%%%%%%%%%%%%%%%%%%%%%%%%%%%%%%%%%%%%%%%%%%%%%%%%%%%%%%%%%%%%%%%%%%%%%%%%%%%%%%%%%%%%%%%%%
\section*{Appendix 2}
\begin{table}[thbp]
    \centering
    \def\hskiplocal{\hskip 0.4cm}
    \begin{tabular}{l@{\hskiplocal}c@{\hskiplocal}c@{\hskiplocal}c@{\hskiplocal}c}
                            & $\relErrorSComponent{xx}$ & $\relErrorSComponent{yy}$ & $\relErrorSComponent{zz}$ & $\relErrorTComponent{\max}$ \\ \midrule
        Plain vanilla PINN        &  3.81        & 5.29         & 3.31        & 7.41                       \\
        Data-enhanced PINN  &  0.15        & 0.17         & 0.16        & 0.18                        \\ \bottomrule
    \end{tabular}
    \caption{Comparison of relative $L_2$ errors (\%) between the predictions and the analytical solution for the plain vanilla PINN and 
    the data-enhanced PINN.}
    \label{tab:comparison_pinn_analytical}
\end{table}
%%%%%%%%%%%%%%%%%%%%%%%%%%%%%%%%%%%%%%%%%%%%%%%%%%%%%%%%%%%%%%%%%%%%%%%%%%%%%%%%%%%%%%%%%%%%%%%%%%%%%%%%%%%%%%%%%%%%%%%%%%%%%%%%%%%%%%%%%%%%%%%%%%%%
\section*{Appendix 3}
\begin{figure}[thbp]
    \def\figsize{0.2}
    \def\hsp{\hspace{-1.cm}}
    \def\hskiplocal{\hskip 0.25cm}
    \centering
    \def\arraystretch{0.2}% 
    \begin{tabular}{cc@{\hskiplocal}cc@{\hskiplocal}cc@{\hskiplocal}cc}
    &\scalebox{1}{Displacement} 
    & 
    &\scalebox{1}{Norm. stress} 
    &
    &\scalebox{1}{Shear stress} 
    &
    &\scalebox{1}{Norm. stress polar}\\[1ex]
    \rownameform{\hspace{-1cm}\scalebox{1}{$x$-direction}}&
    \includegraphics[width=\figsize\linewidth]{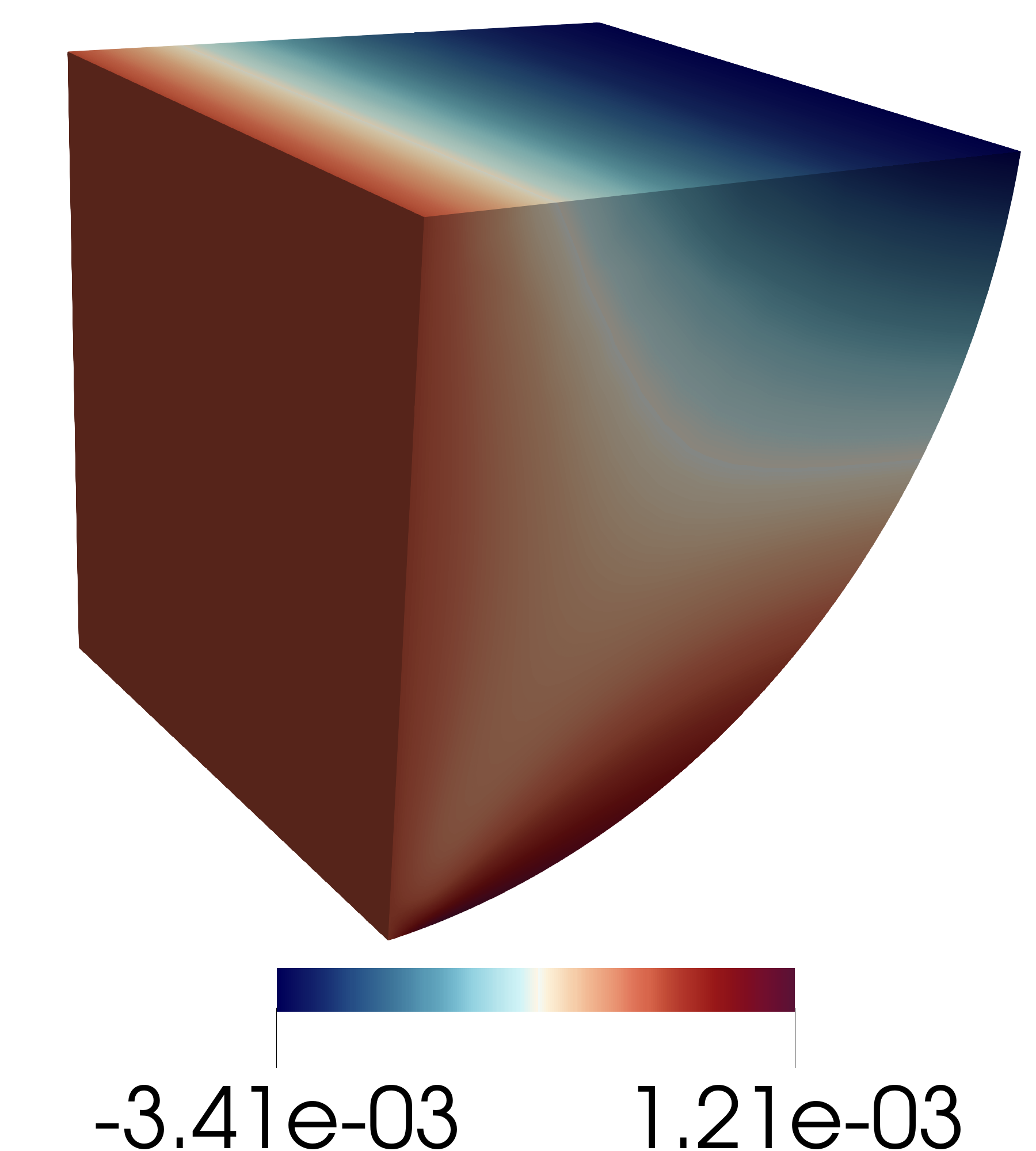}&
    \rownameform{\hspace{-1cm}\scalebox{1}{$xx$-direction}}&
    \includegraphics[width=\figsize\linewidth]{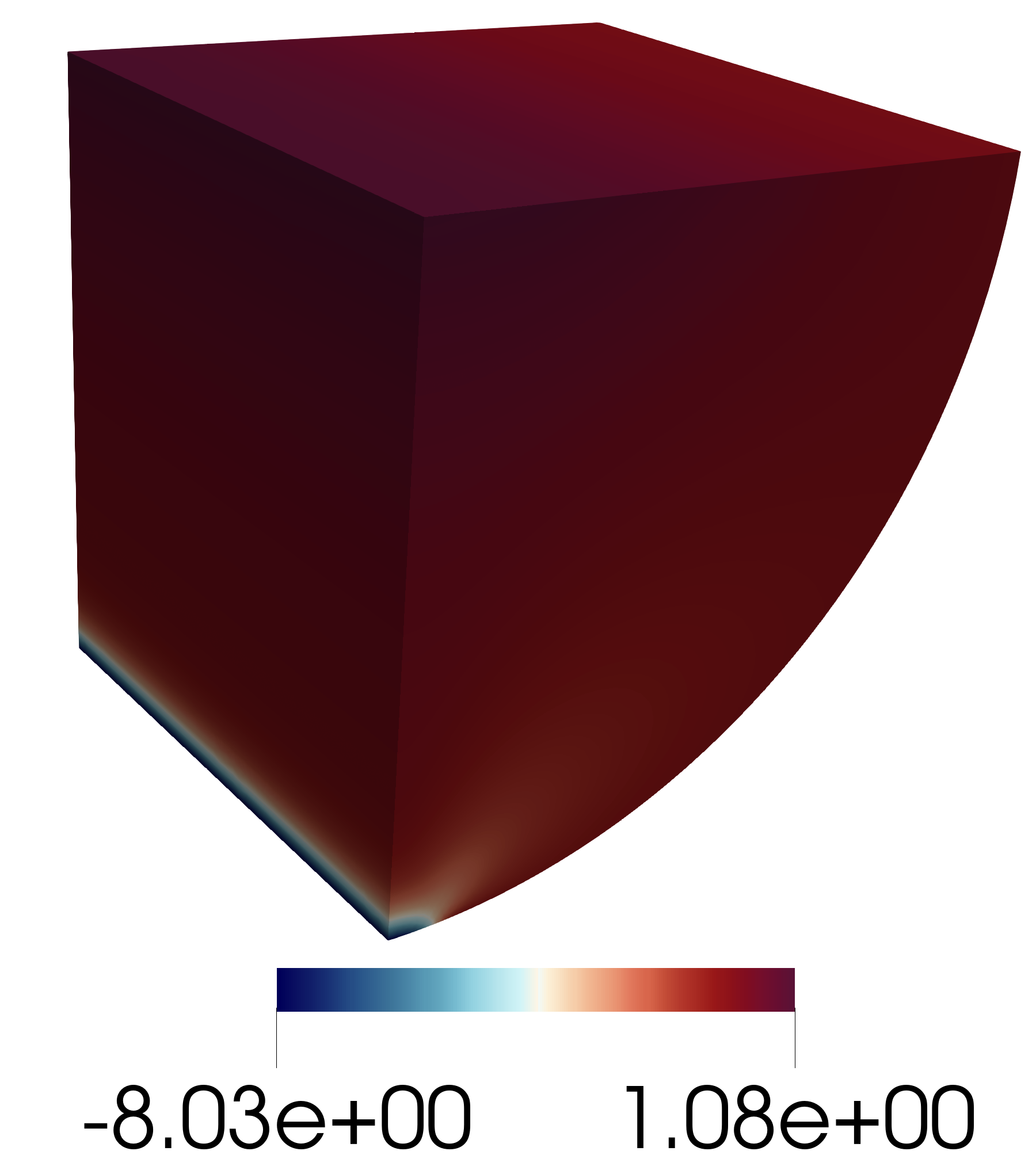}&
    \rownameform{\hspace{-1cm}\scalebox{1}{$xy$-direction}}&
    \includegraphics[width=\figsize\linewidth]{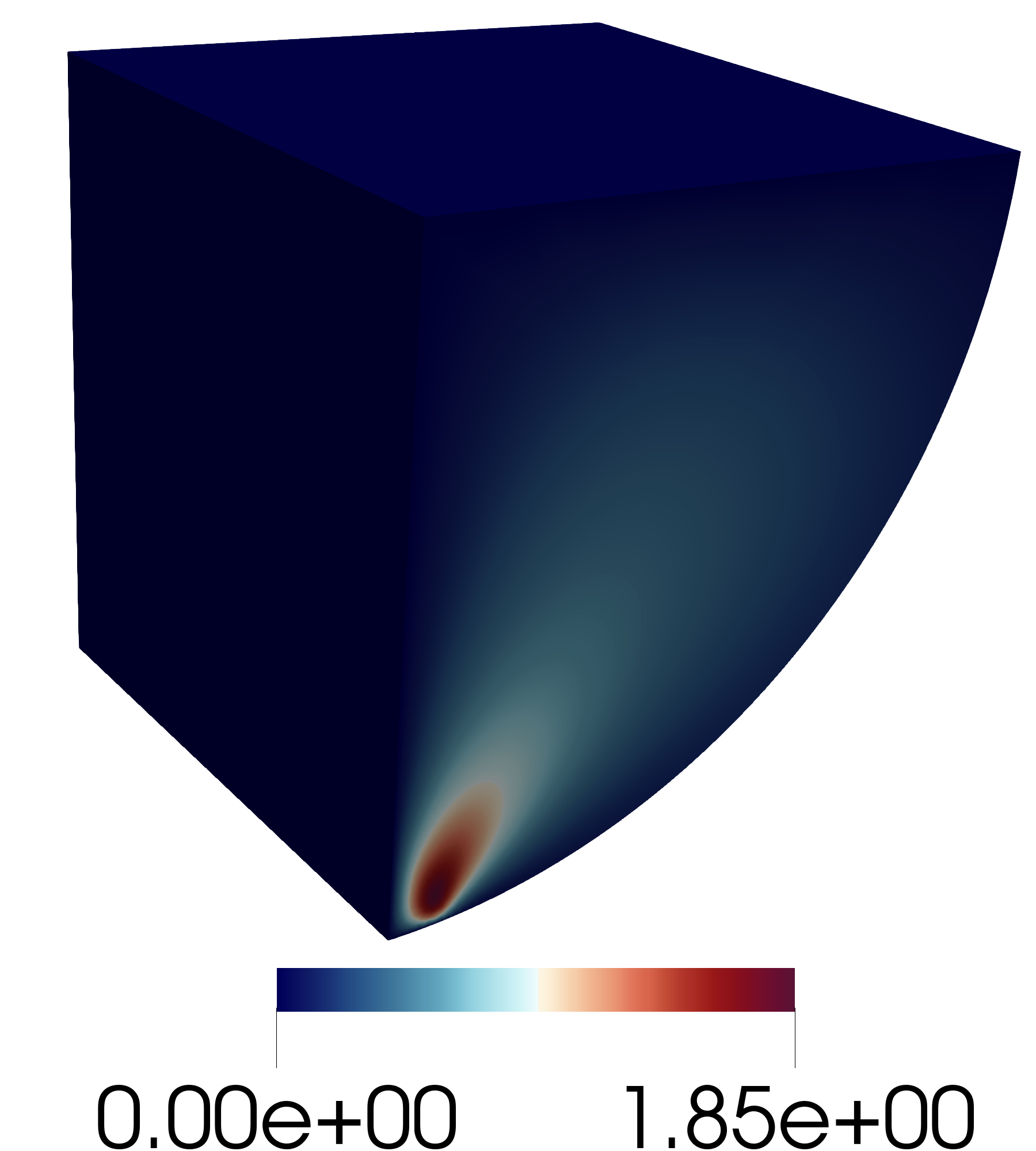}&
    \rownameform{\hspace{-1cm}\scalebox{1}{$rr$-direction}}&
    \includegraphics[width=\figsize\linewidth]{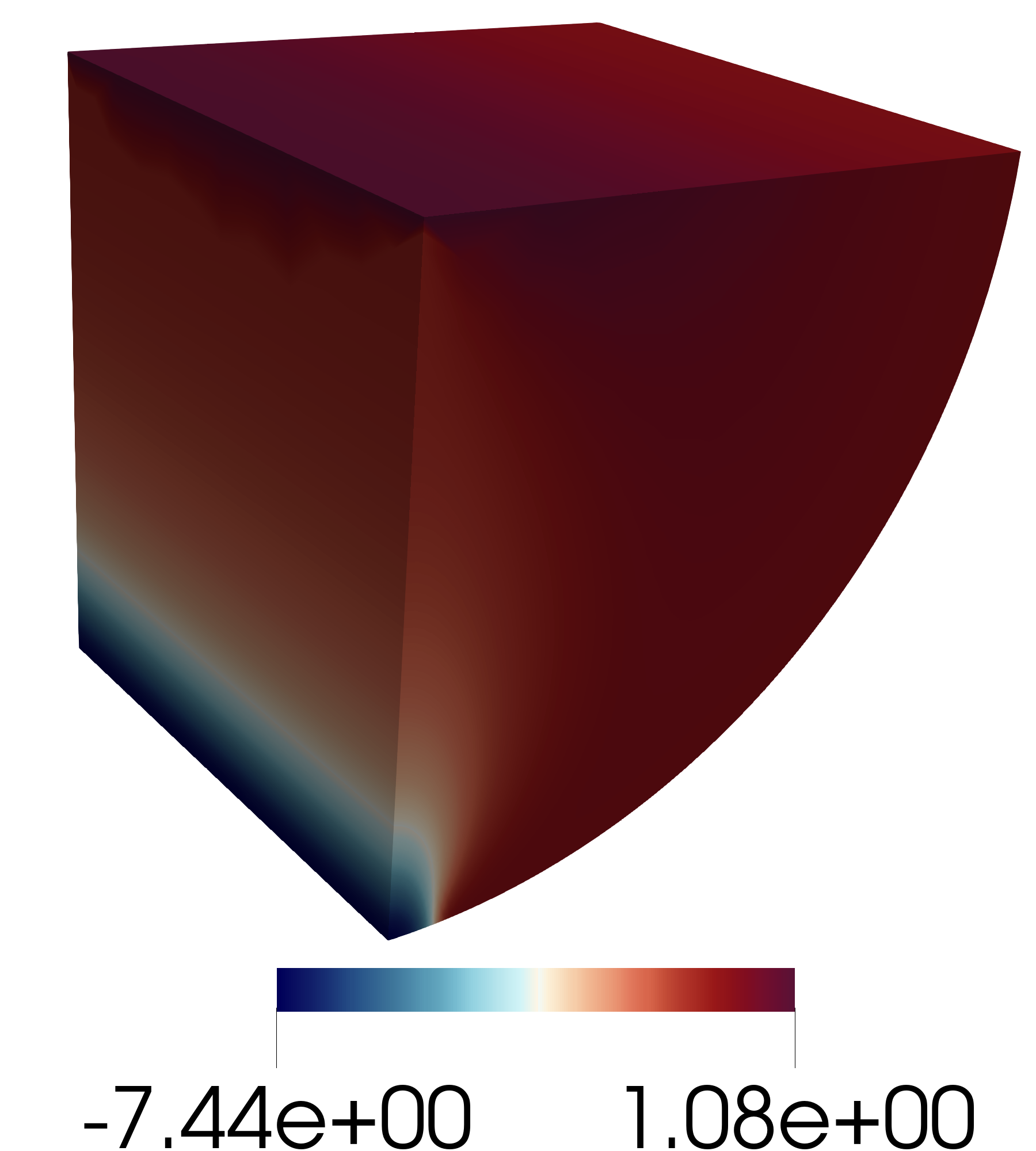}\\[1ex]
    \rownameform{\hspace{-1cm}\scalebox{1}{$y$-direction}}&
    \includegraphics[width=\figsize\linewidth]{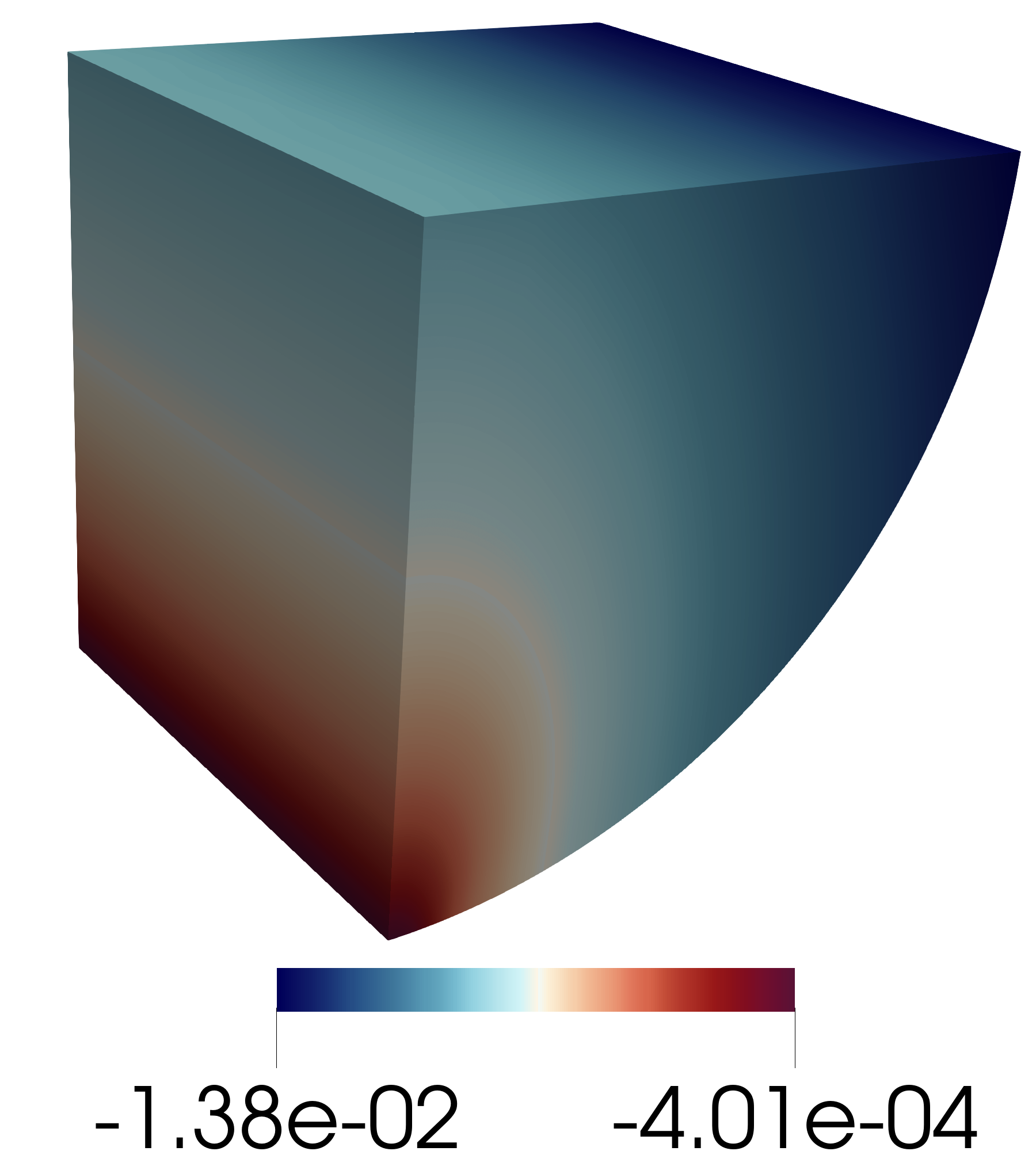}&
    \rownameform{\hspace{-1cm}\scalebox{1}{$yy$-direction}}&
    \includegraphics[width=\figsize\linewidth]{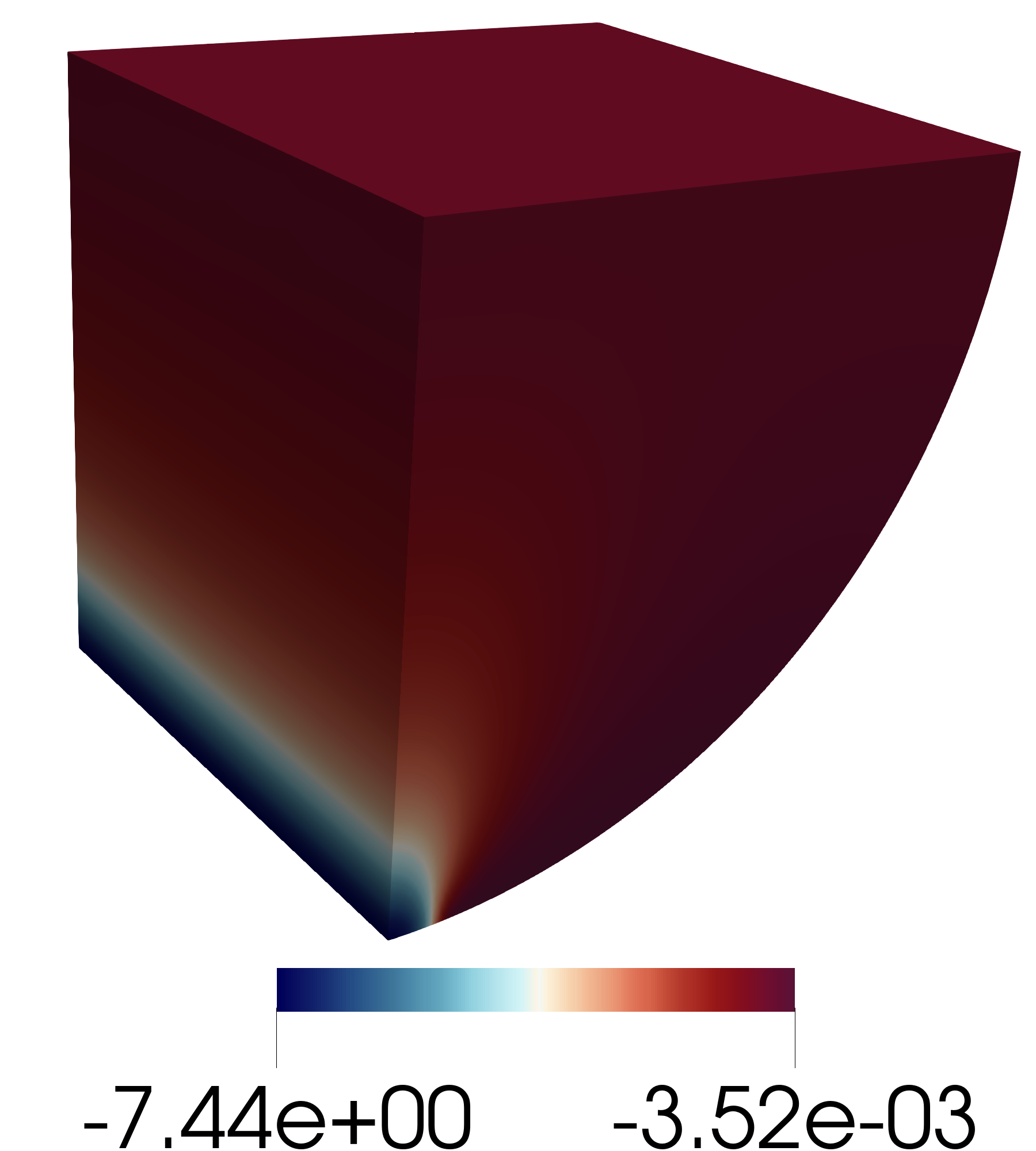}&
    \rownameform{\hspace{-1cm}\scalebox{1}{$yz$-direction}}&
    \includegraphics[width=\figsize\linewidth]{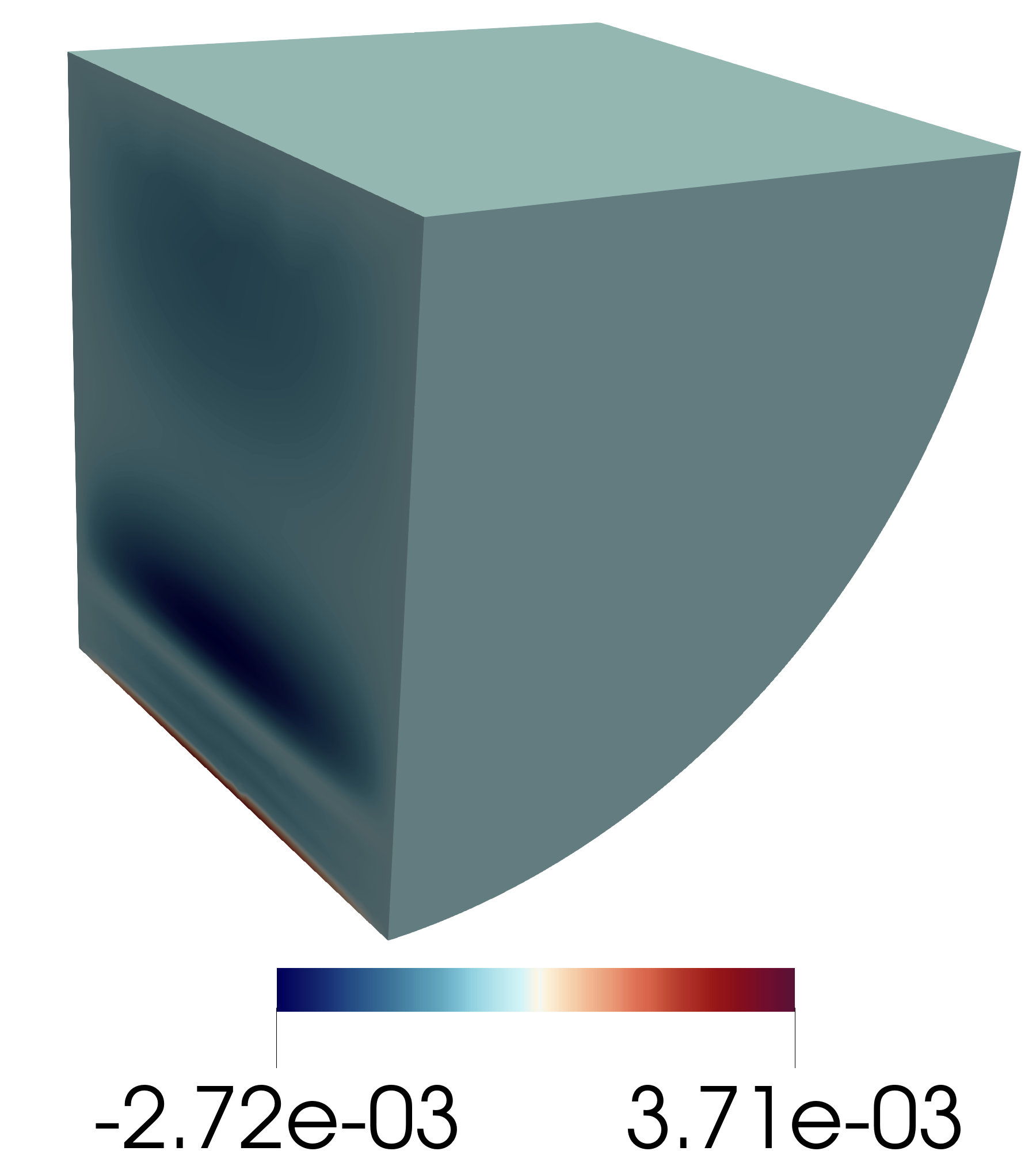}&
    \rownameform{\hspace{-1cm}\scalebox{1}{$\theta \theta$-direction}}&
    \includegraphics[width=\figsize\linewidth]{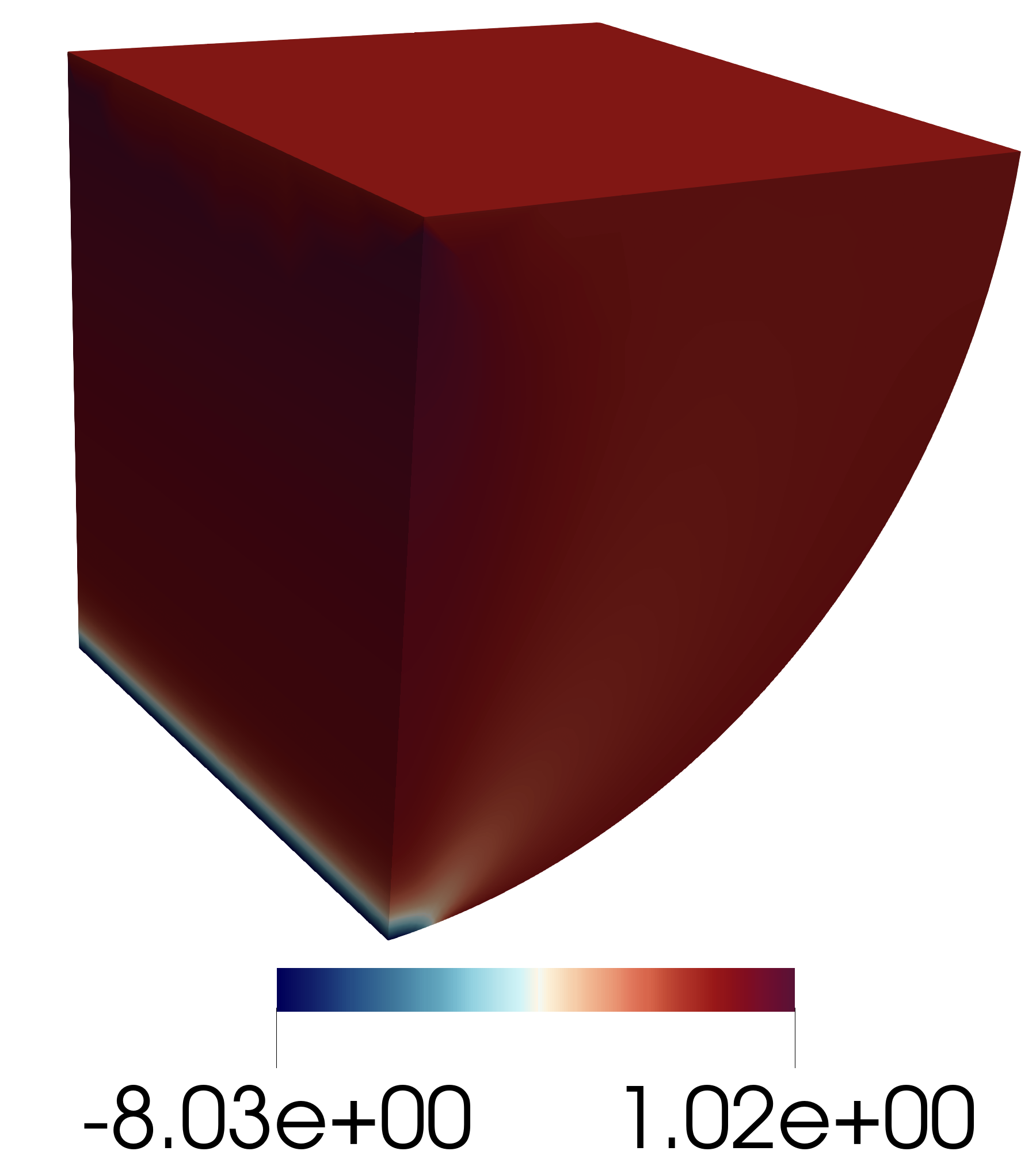}\\[1ex]
    \rownameform{\hspace{-1cm}\scalebox{1}{$z$-direction}}&
    \includegraphics[width=\figsize\linewidth]{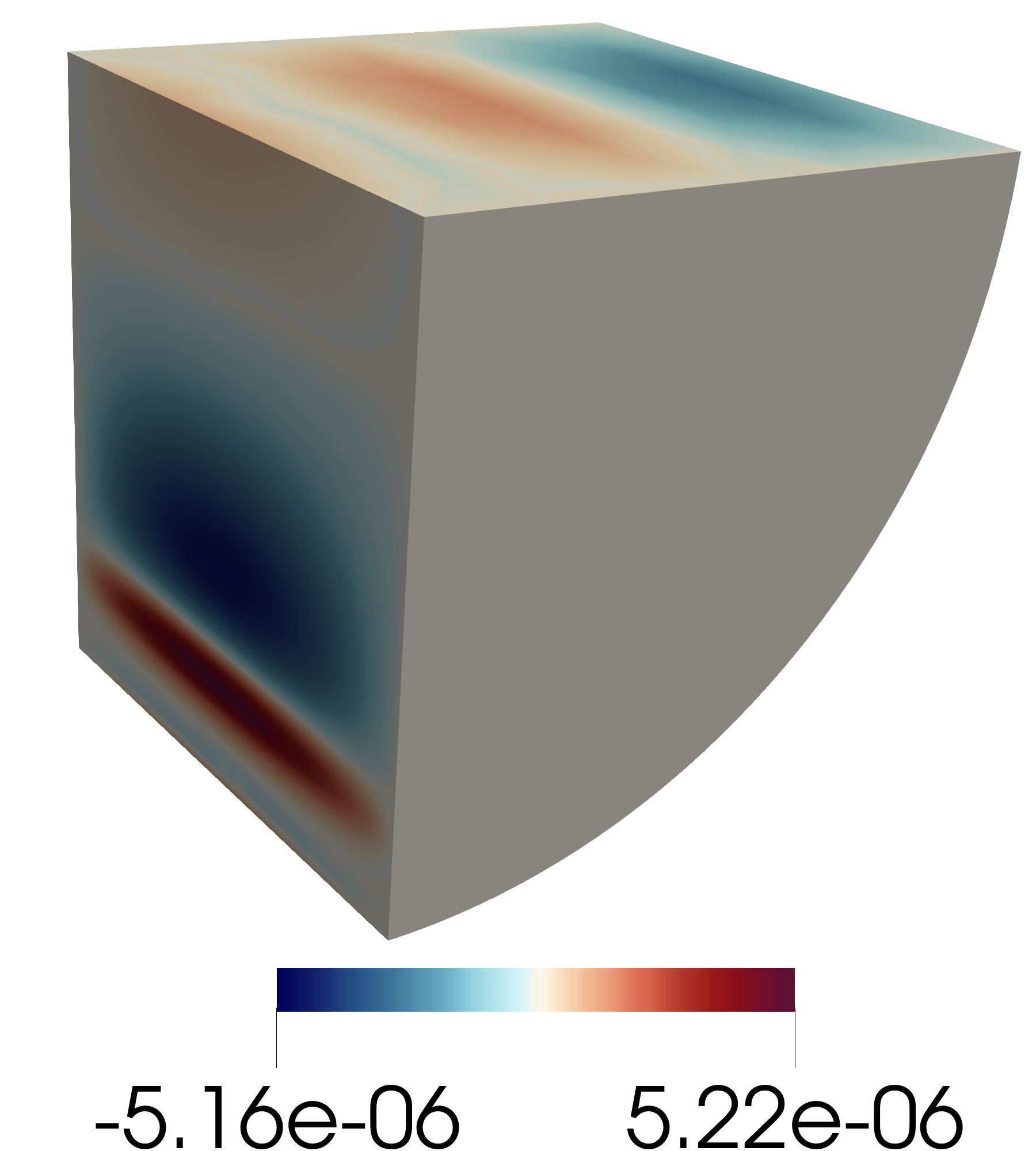}&
    \rownameform{\hspace{-1cm}\scalebox{1}{$zz$-direction}}&
    \includegraphics[width=\figsize\linewidth]{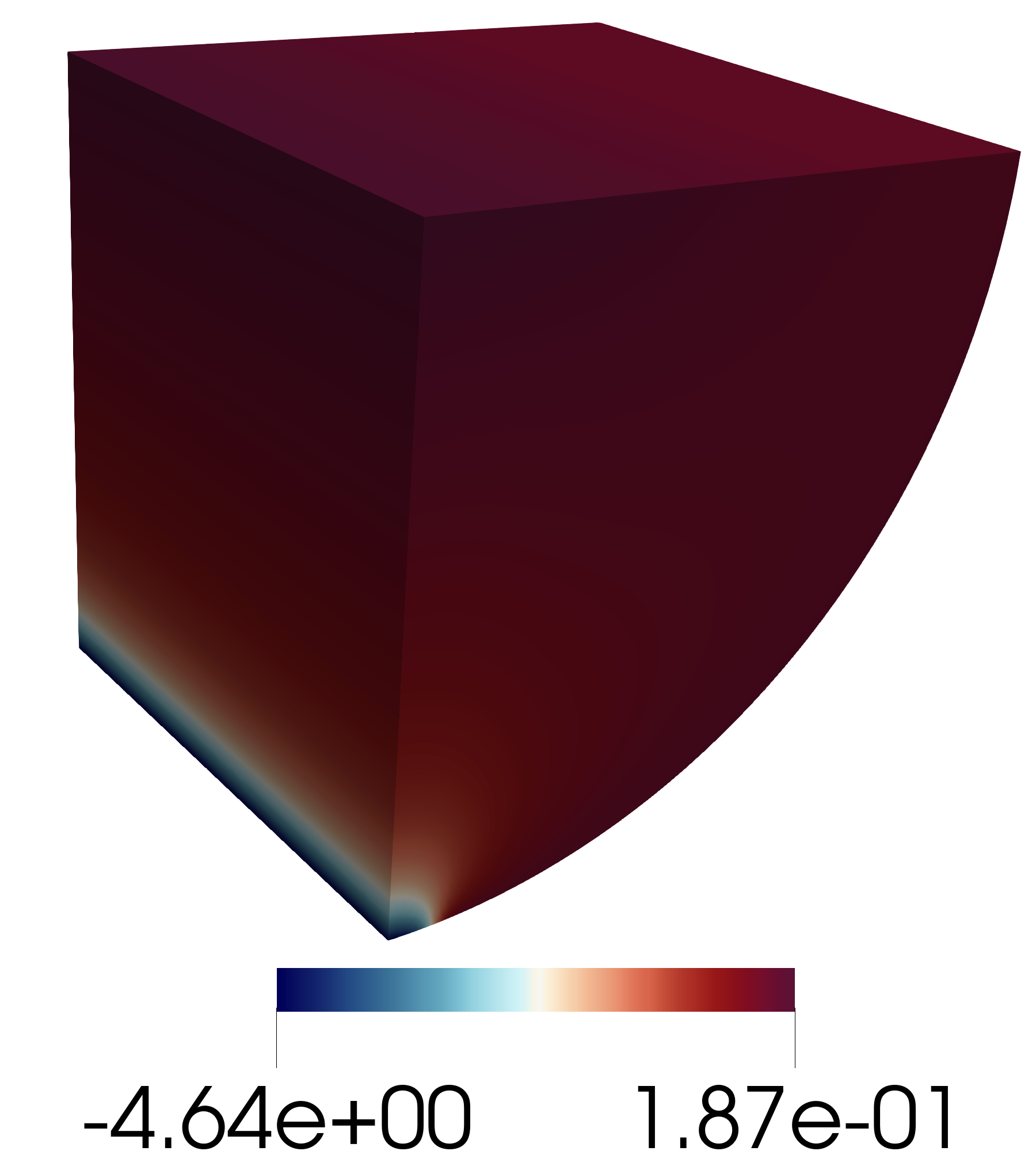}&
    \rownameform{\hspace{-1cm}\scalebox{1}{$xz$-direction}}&
    \includegraphics[width=\figsize\linewidth]{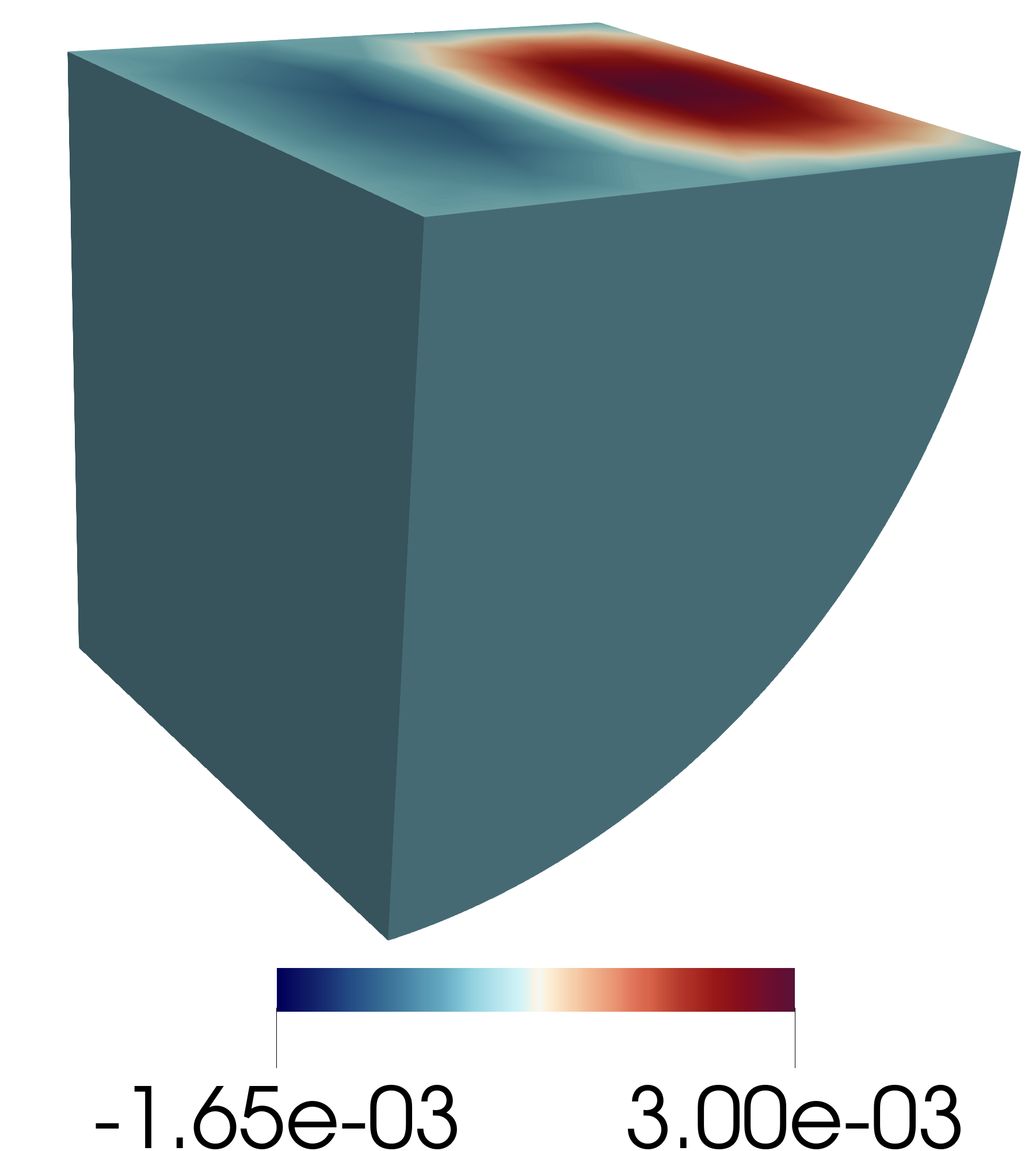}&
    \rownameform{\hspace{-1cm}\scalebox{1}{$zz$-direction}}&
    \includegraphics[width=\figsize\linewidth]{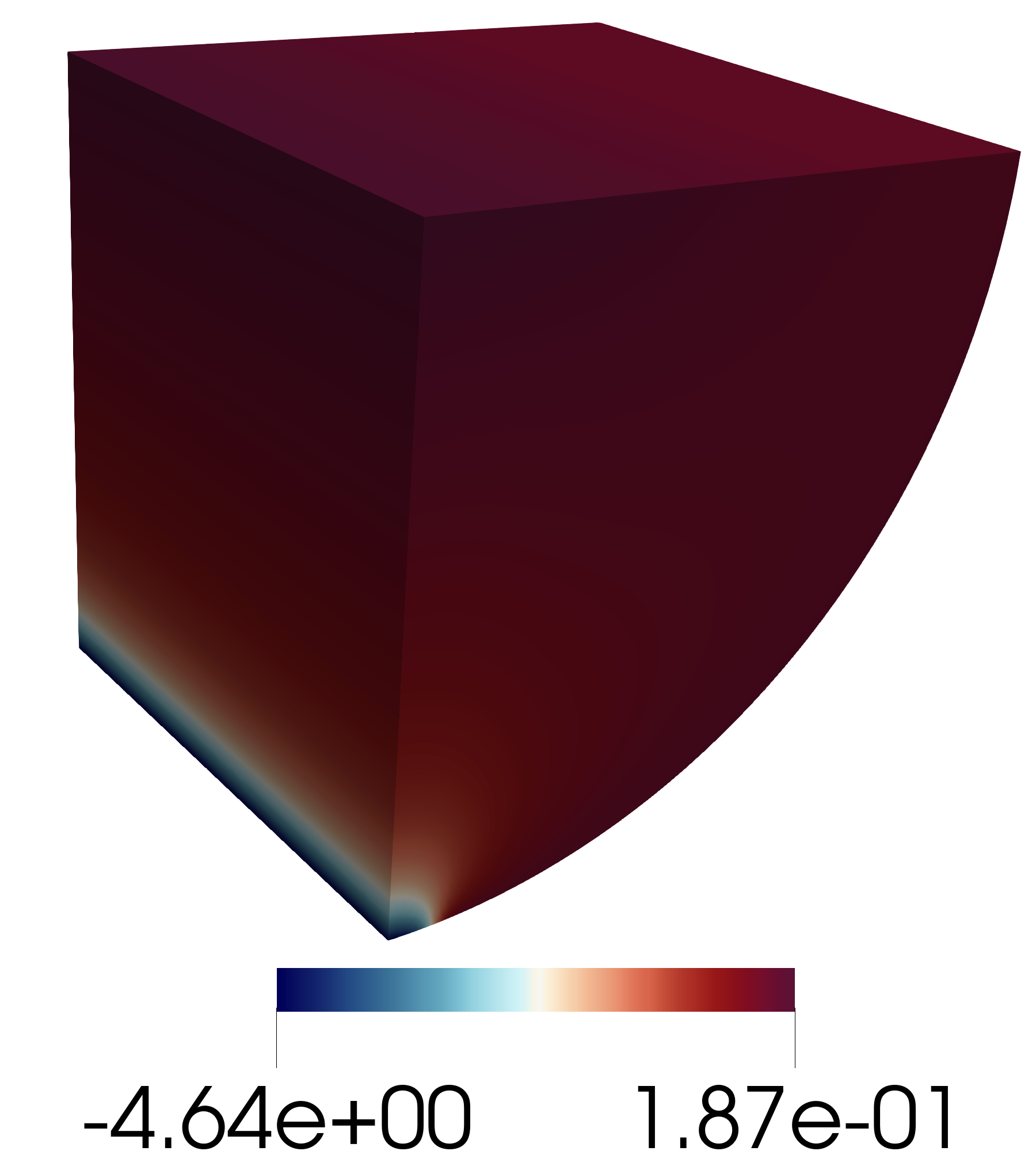}\\%[-1ex]
\end{tabular}
    \caption{The predictions of the plain vanilla PINN for the Hertzian contact example. Polar components are computed based on the cylindrical coordinates.}
    \label{fig:result_hertzian_vanilla}
\end{figure}

%%%%%%%%%%%%%%%%%%%%%%%% referenc.tex %%%%%%%%%%%%%%%%%%%%%%%%%%%%%%
% sample references
% %
% Use this file as a template for your own input.
%
%%%%%%%%%%%%%%%%%%%%%%%% Springer-Verlag %%%%%%%%%%%%%%%%%%%%%%%%%%
%
% BibTeX users please use
% \bibliographystyle{}
% \bibliography{}
%
%\biblstarthook{
% \section{Styling of References}
% References shall be \textit{cited} in the text by number. \footnote{Make sure that all references from the list are cited in the text. Those not cited should be moved to a separate \textit{Further Reading} section or chapter.} The reference list should be arranged in ascending order. 
% The \textit{styling} of references depends on the subject of your book:
% \begin{itemize}
% \item The recommended styles for references in books on \textit{mathematical, physical, statistical and computer sciences} are depicted. 
% \end{itemize}
%}

\end{document}